\documentclass[%
 aip,
 amsmath,amssymb,
 reprint,%
]{revtex4-1}
\usepackage[dvipdfmx]{graphicx}
\usepackage{bmpsize}
\usepackage{calc}
\usepackage{amsmath}
\usepackage{physics}
\usepackage{bm}          %\bm{太字} \mathbb{黒板太字}
\usepackage{mathrsfs}  %\mathcal{筆記体} \mathscr{花文字} \mathfrak{ドイツ文字} \mathsf{サンセリフ}
\usepackage{amssymb}
\usepackage{comment}
\usepackage{cases}
\usepackage{here}
\usepackage{array}
\usepackage{xr}
\usepackage{color}
\usepackage{tabularx}
\newcolumntype{C}{>{\centering\arraybackslash}X}
\newcolumntype{L}{>{\raggedright\arraybackslash}X}
\newcolumntype{R}{>{\raggedleft\arraybackslash}X}
\usepackage{ulem}

\newcommand{\HyRe}{{\mathrm{Re}}}
\newcommand{\HySt}{\mathrm{St}}
\newcommand{\leri}[1]{\left({#1}\right)}
\newcommand{\sgn}[1]{\mathrm{sgn}\leri{{#1}}}

\begin{document}

\title{Vortex phase matching of a self-propelled model of fish with autonomous fin motion
}
\author{Susumu Ito and Nariya Uchida}
\email{uchida@cmpt.phys.tohoku.ac.jp}
\affiliation{Department of Physics, Tohoku University, Sendai, 980-8578, Japan}
\date{\today} 

\begin{abstract}
It has been a long-standing problem how schooling fish 
optimize their motion by exploiting the vortices shed by the others.
A recent experimental study showed that a pair of fish reduce 
energy consumption by matching the phases of their tailbeat
according to their distance.
In order to elucidate the dynamical mechanism by which 
fish control the motion of caudal fins 
via vortex-mediated hydrodynamic interactions, 
we introduce a new model of a self-propelled swimmer 
with an active flapping plate. 
The model incorporates the role of the central pattern generator network
that generates rhythmic but noisy activity of the caudal muscle, 
in addition to hydrodynamic and elastic torques on the fin.
For a solitary fish, the model reproduces 
a linear relation between the swimming speed and tailbeat frequency,
as well as the distributions of the speed, tailbeat amplitude, and frequency.
For a pair of fish, both the distribution function and energy dissipation 
rate exhibit periodic patterns as functions of the front-back distance and  
phase difference of the flapping motion.
We show that a pair of fish spontaneously adjust 
their distance and phase difference 
via hydrodynamic interaction to reduce energy consumption.
\end{abstract}

\maketitle
%%%%%%%%%%%%%%%%%%%%%%%%%%%%%%%%%%%%%%%%%%%%%%%%%%%%%%%%%%%%

\section{Introduction}

Collective behavior of biological units such as insects, birds, mammals and fish 
are ubiquitously found in Nature and have attracted attention for many years~\cite{Conradt2005,Vicsek2012}.
Schooling fish exhibit 
various patterns of collective motion~\cite{Parrish2002,Lopez2012,Terayama2015},
for avoiding predators~\cite{Parrish2002},
foraging food~\cite{Harpaz2020},
and reducing hydrodynamic cost of swimming~\cite{Liao2007}.
They have been studied by
agent-based models that
regard an individual fish as a self-propelled particle 
with positional and orientational degrees of freedom~\cite{
Breder1954, %%% potential
Aoki1982, %%% zoning
Huth1992,Huth1994, %%% zoning
Niwa1994, %%% potential
Couzin2002, %%% zoning 
Hemelrijk2008, %%% gravitiy sensing
Gautrais2012,Calovi2014,%%% topological
Bastien2020,%%% eyesight
Ito2022a, Ito2022b,
Tchieu2012,%%%
Gazzola2016,%%%
Filella2018, %%% topological + hydrodynamic
Deng2021}. %%% predator + hydrodynamic
Interactions between fish are modeled by a potential~\cite{Breder1954,Niwa1994} 
or zones with a blind angle~\cite{Aoki1982,Couzin2002,Huth1992,Huth1994},
while recent works incorporate 
topological interactions~\cite{Gautrais2012,Calovi2014,Filella2018,Deng2021,Ito2022a,Ito2022b},
gravity sensing~\cite{Hemelrijk2008,Ito2022b}, 
and visual information~\cite{Bastien2020}. 
Hydrodynamic interactions 
are taken into account by time-averaged 
dipolar flow~\cite{
Tchieu2012, %%% hydrodynamic dipole 
Gazzola2016, %%% hydrodynamic dipole + leaning 
Filella2018, %%% topological + hydrodynamic
Deng2021%%% predator + hydrodynamic
}.
However, the self-propelled particle models do not describe
the motion of caudal fins,
which is matched to the vortex flow to reduce muscle activity~\cite{Liao2007}.

An undulating caudal fin sheds a reverse K\'arm\'an vortex street, 
whose vorticity has an opposite sign compared to 
a K\'arm\'an vortex street~\cite{Lauder2002,Akanyeti2017,Wise2018}.
It has long been hypothesized that schooling fish  
exploit the reverse K\'arm\'an vortex 
to reduce energy consumption~\cite{Breder1965,Weihs1973}.
Weihs proposed that fish form a two-dimensional diamond lattice structure to reduce hydrodynamical drag force caused by the vortices~\cite{Weihs1973}.
Later observations on some species of fish, 
% e.g.
% saithe (\textit{Pollachius virens}),
% herring (\textit{Clupea harengus}),
% cod (\textit{Gadus morhua}), and
% golden grey mullet (\textit{Liza aurata}),
however, demonstrated that a fish school does not preserve 
a specific lattice structure~\cite{Partridge1979,Marras2014}.
More recent experiments revealed that
red nose tetra (\textit{Hemigrammus bleheri})
synchronize their tailbeat with the nearest neighbors
at high swimming speed~\cite{Ashraf2016,Ashraf2017}.
Li \textit{et al.}~\cite{Li2020} found a linear relation between 
the phase difference of the tailbeat and the front-back distance between 
a pair of goldfish (\textit{Carassius auratus}),
which shows that the motion of the caudal fins 
are regulated by the periodicity of the vortices
and weakly phase-locked at any short distance. 
Using robotic fish, they also found 
that energy consumption is reduced by matching 
the phases of the vortices generated by two fish~\cite{Li2020}.
These results indicate
the necessity of a theoretical model that describes 
autonomous motion and phase locking of caudal fins.
Since hydrodynamic synchronization does not necessarily 
lead to minimization of energy dissipation~\cite{Elfring2009,Liao2021},
the dynamical mechanism exploited by fish to reduce 
energy consumption is highly nontrivial.

%%% 
% Breder1954,Aoki1982,Huth1992,Huth1994,Niwa1994,
% Gautrais2012,Calovi2014,Bastien2020,Ito2022a,Ito2022b,
%%% 自己駆動、流体なし
% Tchieu2012,Gazzola2016,Filella2018,Deng2021,
%%%　自己駆動、流体あり、渦なし
%%%%%%%%%%%%%%%%%%%%%%%%%%%%%%%%%%%%%%%%%%%%%%%
% Dewey2014,
%%% 翼形，実験と解析，2 体横列，回転振動，位置と位相固定，位相差変えて反位相で遊泳効率最大
% Boschitsch2014,
%%% 翼形，主に実験，2 体縦列，回転振動，位置と位相固定，翼形，位相差と距離を変えて遊泳効率に周期性
% Becker2015,
%%% 翼形，実験（2 体回転）と CFD （１次元隊列），位置と位相固定，同位相と逆位相でパワー比較
% Newbolt2019,
%%% 翼形，実験（2体回転），上下振動，位相固定，位置は自発変化，位相差を変えて安定な距離
% Ramananarivo 2016, 
%%% 翼形，実験と解析（2体），上下振動，位相固定，位置は自発変化，安定な距離に離散性
% Oza2019}.
%%% 解析，1次元 & 2 次元隊列，上下振動，位相固定，位置は自発変化，安定な距離に離散性
%%%%%%%%%%%%%%%%%%%%%%%%%%%%%%%%%%%%%%%%%%%%%%%%
% Hemelrijk2015,
%%%　fish-shape, CFD, 1D & 2D 格子隊列，距離と隊形変えて遊泳効率や速度
% Daghooghi2015,
%%%　fish-shape, CFD, 2D 格子隊列（長方形），距離変えて遊泳効率
% Maertens2017,
%%%  fish-shape, CFD, 2体，泳法の比較，位相差と振動数変えて遊泳効率
% Li2019, 
%%% fish-shape, CFD, 2体，横列と縦列の比較
%%%%%%%%%%%%%%%%%%%%%%%%%%%%%%%%%%%%%%%%%%%%%%%%
% Zhu2014,
%%%  弾性フィン，CFD，2体，位相固定，位置は自発変化，安定な距離に離散性
% Park2018,
%%% 弾性フィン，CFD, 1D & 2D （1-4 個）隊列，位相固定，位置は自発変化，安定な距離に離散性
% Peng2018,
%%% 弾性フィン，CFD, 1D隊列，位相は固定，位置は自発変化，安定な距離に離散性
%
%Self-propelled particle models usually do not introduce
%fluid~\cite{Breder1954,Aoki1982,Huth1992,Huth1994,Niwa1994,
%Gautrais2012,Calovi2014,Bastien2020,Ito2022a,Ito2022b},
%or can treat fluid flows without 
%the vortex street~\cite{Tchieu2012,Gazzola2016,Filella2018,Deng2021}:
%the Reynolds number is laminar region ($\HyRe\sim\mathcal{O}(10^1)$--$\mathcal{O}(10^3)$).

Previous models of vortex-mediated interaction between fish use  
(i) flapping~\cite{Dewey2014,Boschitsch2014} or
heaving~\cite{Becker2015,Ramananarivo2016,Newbolt2019,Oza2019}
airfoils,
(ii) elastic filaments~\cite{Zhu2014,Park2018,Peng2018},
or
(iii) deformable fish-shaped swimmers~\cite{Hemelrijk2015,Daghooghi2015,Maertens2017,Li2019,XLi2021,Pan2022,Kelly2023,Lin2023}.
The type (i) models allow analytical treatment by Joukowski transformation~\cite{Dewey2014,Ramananarivo2016,Oza2019}  
and direct comparison with experiments~\cite{Dewey2014,Boschitsch2014,Becker2015}.
Computational fluid dynamics simulations are employed 
for the type (ii) and (iii) models.
The type (ii) models treat hydroelastic deformation of the filament  in two dimensions
induced by prescribed oscillation of the filament head. 
%plunging motion. %(vertical oscillation of the filament head).
% and 
%assumed a constant bending rigidity,
%while it was varied along the filament in a one-body study~\cite{Gazzola2015}.
The type (iii) models 
prescribe undulatory motion of a three-dimensional fish-shaped body.
%(iii) 
%is prescribed and mimics carangiform or subcarangiform,
% which the majority of fish adopt~\cite{Sfakiotakis1999,Lauder2005}.
%which can handle turbulent flow at the
%typical Reynolds number of $10^4$--$10^5$ ~\cite{Gazzola2014,Li2020}.
%$\HyRe\sim\mathcal{O}(10^4)$--$\mathcal{O}(10^5)$
%%% {Re の値必要？Gazzola2014, Fig.2: 魚の Re=$10^4-10^8$. Li2021, Fig. 5: CFD の Re=$1-10^3$.} 
%%% \NU{Gazzola2015 は 1体だが紹介しておく．}
% e.g. mullet (\textit{Chelon labrosus})~\cite{Hemelrijk2015},
% atlantic mackerels (\textit{Scomber scombrus})~\cite{Daghooghi2015},
% giant danio(\textit{Devario aequipinnatus})~\cite{Maertens2017},
% and red nose tetra~\cite{Li2019}.
%%% 魚種の言及は必要？
% The fish-shaped swimmers designate the relative distance and 
% the phase difference of neighbors,
% which do not change spontaneously.
% In the elastic filaments models, 
% the relative distance can change with time evolution,
% but they study in the case of specific phase differences (synchronization and antisynchronization).
%% 
%These studies discuss 
Optimal swimming patterns are discussed 
for two bodies~\cite{Dewey2014,Boschitsch2014,Becker2015,Ramananarivo2016,Newbolt2019,Maertens2017,Li2019,Zhu2014,Peng2018,Lin2023}
and a lattice~\cite{Hemelrijk2015,Daghooghi2015,Oza2019,Park2018,XLi2021,Pan2022,Kelly2023} of fish.
Some studies incorporate temporal evolution of the distance between swimmers~\cite{Ramananarivo2016,Newbolt2019,Zhu2014,Park2018,Peng2018},
but the phases of the oscillation or undulation are prescribed in all the models. 
%%%
% 2体：特定の位相で安定
%格子：特定の格子だとエネルギーが下がる．Oza, Park, Peng
% ３種類のモデルをまとめて位相固定のものと位相、位置両方固定のものに分ける．
%%%

In this paper, we introduce a minimal but integrated model of fish
that self-propels by autonomous fin motion
and interacts via a reverse K\'arm\'an vortex street.
By incorporating physiological noises 
in the signals transmitted to the caudal muscle,
the model describes spontaneous time evolution of 
the phase of the caudal fin, which is modeled by an active flapping plate.
The noises are essential in reproducing
the distribution of the swimming speed of a solitary swimmer
and the vortex-mediated correlation between the distance and phases of two swimmers
that are experimentally observed~\cite{Li2020}.

This paper is organized as follows.
In Sect. II, we construct the model incorporating 
hydrodynamical forces,
elasticity of the caudal fin, 
and an active physiological noises.
In Sect. III, to test the validity of the model,
we analyze the properties of solitary swimming and compare them with experimental results.
In Sect. IV, we consider a pair of swimmers and show that
the correlation between their distance and phase difference
reflects the periodicity of the vortices.
We also show that the fish tend to be distributed
at short distances where they adjust their phase difference to 
reduce energy dissipation.
We discuss the results in comparison with previous studies in Sect. V.

%%%%%%%%%%%%%%%%%%%%%%%%%%%%%%%%%%%%%%%%%%%%%%%%%%%%%%%%%%%%
\section{Flapping plate model}
%//////////////////////////////////////////////////////////////////////////////////////////////////////////////////////////%
\subsection{Equations of motion and forces}
\begin{figure}[!b]
\centering
\includegraphics[width=\linewidth]{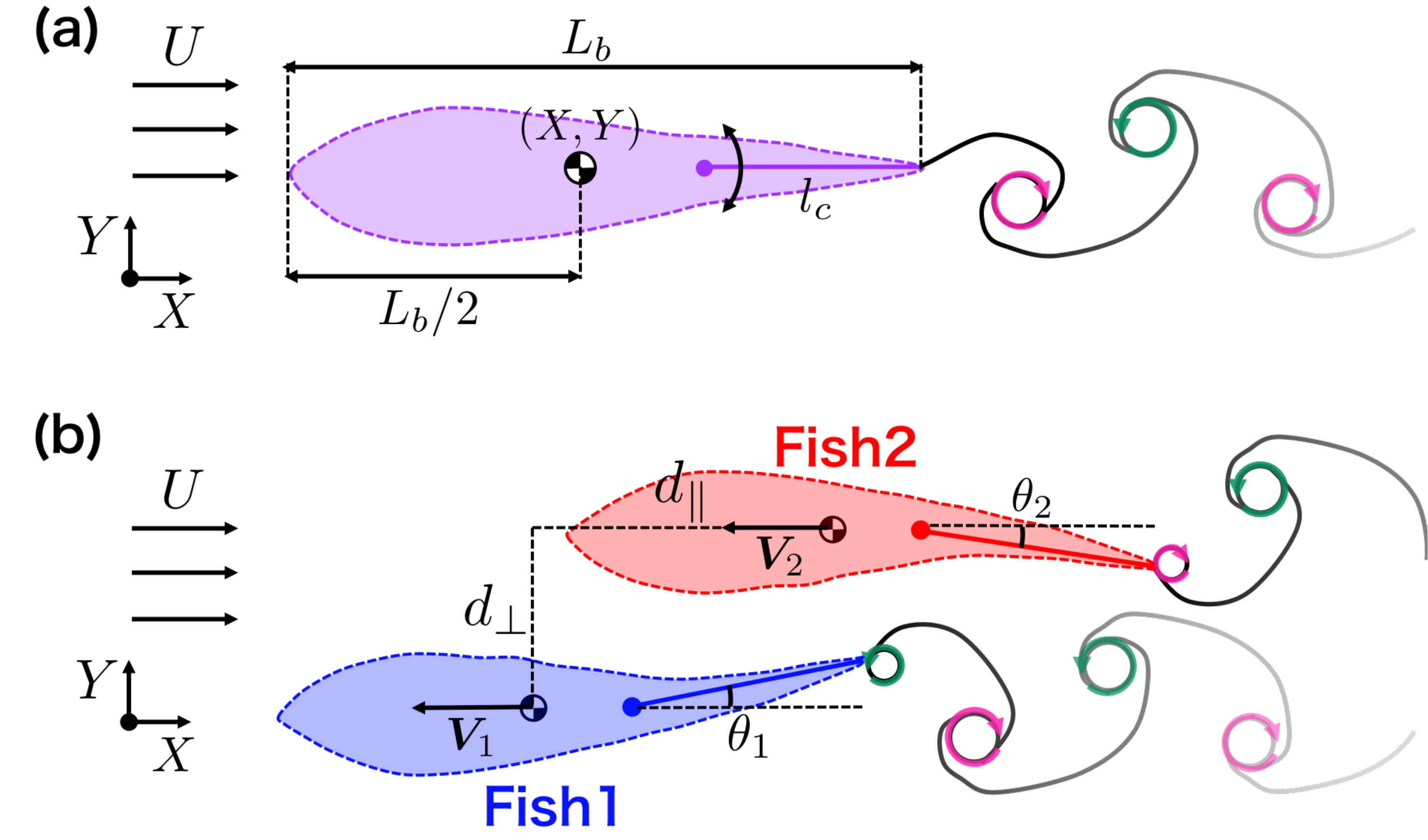}
%bb=0 0 360 252
\caption{
Schematic illustration of (a) a solitary swimmer and (b) a pair of swimmers. 
The swimmers are placed in a uniform background flow of speed $U$ along the $X$-direction.
In (a), the center of the swimmer is at $(X,Y,0)$. 
The right top inset shows the swimmer viewed from the side.
The configuration of a pair of swimmers is specified by
the longitudinal distance $d_\parallel=|X_1- X_2|$, the lateral distance $d_\perp=|Y_1 - Y_2|$
and the angles $\theta_1$, $\theta_2$ of the flapping plates. 
(In the figure, $\theta_1 >0$ and $\theta_2<0$).
The vortices shedded by the swimmers are 
carried away by the background flow and diminish. 
}
\label{schimaticswimmer}
\end{figure}

The modes of fish locomotion are classified into 
anguilliform, 
%(undulating motion of the most of body),
subcarangiform, carangiform, thunniform, and
ostraciiform, 
%(which is oscillating motion with the tip of the caudal fin)
in the order of the fraction of oscillating 
body sections~\cite{Sfakiotakis1999,Lauder2005}.
A majority of fish species adopt subcarangiform or carangiform, 
which are characterized by undulating motion
of 30 to 50 percents of the body including the caudal fin~\cite{Sfakiotakis1999,Lauder2005}.
It has been previously studied by a
two-hinge flapping airfoil model, which consists of
a massless rod corresponding to the caudal peduncle and
an airfoil corresponding to the caudal fin~\cite{Azuma1992,Nagai1996,Hirayama2000}.
One end of the rod is connected to the airfoil by a hinge and
the other end is anchored  to an immovable point by the second hinge.
%\Delete{Using the two-dimensional model, the previous works analyzed}
The two-hinge model was numerically solved to analyze
the optimal phase delay between the oscillations of 
the rod and airfoil that brings about
maximum thrust efficiency~\cite{Nagai1996,Hirayama2000}.
On the other hand, 
a simpler single-hinge model enables it to
analytically calculate various swimming characteristics such as 
the thrust, power and swimming efficiency~\cite{Azuma1992}.
Although a more sophisticated model 
%\Delete{of locomotion gait is composed of infinitesimal elastic segments} 
uses an elastic filament
with varying bending stiffness~\cite{Gazzola2015},
the flapping airfoil models have the merit that
the hydrodynamic forces are easier to analyze 
and simulations are computationally less costly.
In order to include the physiological mechanism
and 
vortex-mediated hydrodynamic interactions,
we adopt a single-hinge model as the minimal base model.
We summarize the variables, functions, and parameters of our model in Table~\ref{notationTable}.

We assume that the swimmer consists of 
a streamlined body and a rectangular rigid plate.
The characteristic sizes of the swimmer are 
the total length $L_b$ of the swimmer (which we will call the body length)
including the plate in a straight configuration, 
and the height $H_b$ of the body and plate
and the length $l_c$ of the flapping plate.

By assumption, the center of the fish is constrained 
in the horizontal plane $Z=0$, and its in-plane coordinates 
are denoted by $(X,Y)$.
The fish is placed in a uniform background flow of 
speed $U\geq0$ along the $X$-axis
and swims toward the negative $X$-direction,
as shown in Fig.~\ref{schimaticswimmer}(a).

\begin{table*}[!t]
\begin{center}
\caption{
The  list of the variables, functions, and parameters of our model.
See also Appendix.~\ref{AppendExperi} for the parameter values
and their correspondence with the experiments.
}
\centering
\includegraphics[width=\linewidth]{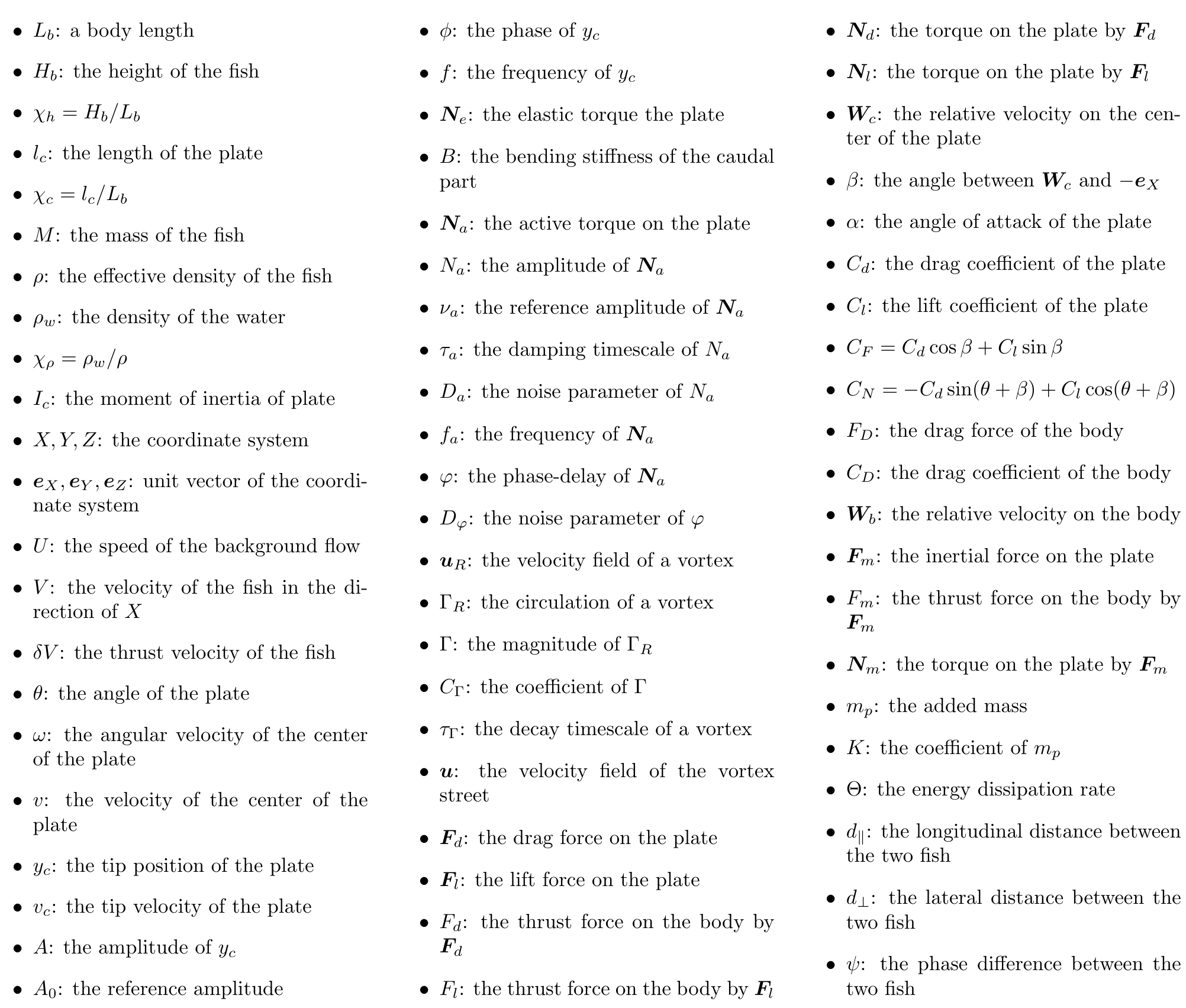}
\label{notationTable}
\end{center}
\end{table*}

Here we derive the equations of motion for a pair of swimmers
(see Fig.~\ref{schimaticswimmer}(b)).
The center position $(X_i,Y_i)$ of the $i$-th swimmer ($i=1,2$)  
moves along the $X$-axis with the velocity $\bm{V}_i= V_i\bm{e}_X$.
(We define the unit vectors $\bm{e}_X$ and $\bm{e}_Y$ and $\bm{e}_Z$ 
along the $X$- and $Y$-axis, and $Z$-axis, respectively.)
Note that negative values of $V_i \equiv \mathrm{d}X_i/\mathrm{d} t$ corresponds to 
forward propulsion of swimmer by convention. 
We also define the deflection angle $\theta_i$ of the flapping plate 
measured anti-clockwise from the $X$-axis,
and the angular velocity $\omega_i=\mathrm{d}\theta_i/\mathrm{d}t$.
%Viewed from positive direction of $Z$-axis$, 
%\theta_i>0$ when a plate to the right
%(for example, $\theta_1>0,~\theta_2<0$ in the case of Fig~\ref{schimaticswimmer}(b)).
The equations of motion for the $i$-th swimmer read
\begin{equation}
\label{EoMF}
M\dv{V_i}{t}=F_{d,i}+F_{l,i}+F_{m,i}+F_{D,i},
\end{equation}
and
\begin{equation}
\label{EoMN}
I_c\dv{\omega_i}{t}\bm{e}_Z=\bm{N}_{e,i}+\bm{N}_{a,i}+\bm{N}_{d,i}+\bm{N}_{l,i}+\bm{N}_{m,i},
\end{equation}
where $M$ is the body mass and $I_c$ is the moment of inertia of the flapping plate.
For the body mass,
we use the empirical formula 
that does not depend on the fish species~\cite{Jones1999}:
\begin{equation}
\label{mass}
M=\rho L_b^2H_b.
\end{equation}
Here, $\rho$ is the effective density of the body
and is smaller than the actual body density 
which approximately equals to the density of water $\rho_w$ ($\rho<\rho_w$).
We assume that 
the flapping plate has 
the same density per unit length $M/L_b$ as the body.
Its moment of inertia is thus given by
\begin{equation}
\label{Ic}
I_c=\frac{M}{L_b}\int_0^{l_c}\dd ll^2=\frac{Ml_c^3}{3L_b}=\frac{\rho}{3}L_bH_bl_c^3.
\end{equation}

In Eqs.(\ref{EoMF},\ref{EoMN}),
$F_{d,i}$, $F_{l,i}$, $F_{m,i}$, and $F_{D,i}$ are the thrust forces exerted on the body,
and 
$\bm{N}_{e,i}$, $\bm{N}_{a,i}$, 
$\bm{N}_{d,i}$, $\bm{N}_{l,i}$, and $\bm{N}_{m,i}$
are the torques exerted on the plate,
of which 
$\bm{N}_{e,i}$ is the passive elastic torque 
and $\bm{N}_{a,i}$ is the active physiological torque 
generated by the caudal muscles.
The other components of the torques 
and all the thrust forces  
are of hydrodynamic origins.
The non-steady hydrodynamic forces acting on an oscillating body
are complex due to turbulent flow
at $\HyRe\sim\mathcal{O}(10^4)$--$\mathcal{O}(10^5)$~\cite{Landau1987}.
However, the amplitude of the tail tip motion is $A_0\sim0.1L_b$
for various species and locomotion gaits
of fish~\cite{Bainbridge1958,Hunter1971,Webb1984,Akanyeti2017,Li2021,NoteA0},
and the small amplitude
allows us to adopt a quasi-steady
approximation~\cite{Nagai1996,Hirayama2000,Gazzola2015},
because oscillation with a small amplitude causes a potential flow~\cite{Landau1987}.
In the quasi-steady approximation, 
we calculate the drag and lift forces from those in steady flow~\cite{Taylor1952},
and the inertial force by the added mass in potential flow~\cite{Lighthill1960}.
The expressions for the thrust forces and torques 
in Eqs.~(\ref{EoMF},\ref{EoMN}) will be derived in the following subsections.

%//////////////////////////////////////////////////////////////////////////////////////////////////////////////////////////%
\subsection{Passive elastic and active physiological torques}

First, we formulate the non-hydrodynamic 
torques $\bm{N}_{e,i}$ and $\bm{N}_{a,i}$ in Eq.~(\ref{EoMN}).
These torques are originated from internal forces that cancel out
over the entire body and
do not contribute to the thrust force in Eq.~(\ref{EoMF})~\cite{Gazzola2015}.

The passive elastic torque
$\bm{N}_{e,i}$ is proportional to the curvature of the caudal fin~\cite{Gazzola2015,Landau1986}.
We approximate the curvature as $\sin\theta_i/l_c$
because the deflection angle $\theta_i$ is small;
the transverse displacement of a plate tip is $l_c\sin\theta_i$
and the second-order spatial derivatives introduces the factor $l_c^{-2}$.
We obtain, therefore,
\begin{equation}
\label{Ne}
\bm{N}_{e,i}=-\frac{B}{l_c}\sin\theta_i\bm{e}_Z,
\end{equation}
where $B$ is bending stiffness of the caudal part 
estimated as 
the product of the Young's modulus and
the moment of inertia per area of a dead fish~\cite{McHenry1995}.

The active physiological torque 
$\bm{N}_{a,i}$ mimics the roll of the central pattern generator network (CPG) 
that generate rhythmic activity of the caudal muscle
without sensory input~\cite{Grillner2006,Song2020}.
Although there are models of the CPG 
which treat the neural circuits in detail~\cite{Ekeberg1999,Matsuoka2011},
we use sinusoidal signals for
simplicity~\cite{Gazzola2015}.
Importantly, we model the physiological noises 
in the signals transmitted from the CPG to the caudal muscle and 
observed by electromyography~\cite{Schwalbe2019}:
the electromyography shows that
there are fluctuations in the amplitude and the duration between the signals.
These noises cause spontaneous changes in
the amplitude and phase of the tailbeat.
We formulate the torque as 
\begin{equation}
\label{Na}
\bm{N}_{a,i}=N_{a,i}(t)\sin(2\pi f_at+\varphi_{i}(t))\bm{e}_Z,
\end{equation}
where $f_a$ is the frequency of the CPG signals
which we call the active frequency.
The amplitude $N_{a,i}$ and the phase-delay $\varphi_{i}$ 
are controlled by an Ornstein-Uhlenbeck and Wiener process, respectively,
and obey the stochastic differential equations
\begin{equation}
\label{Nanoise}
\dv{N_{a,i}}{t}=\frac{1}{\tau_a}(\nu_a-N_{a,i})+\eta_{a,i},
\end{equation}
\begin{equation}
\label{phinoise}
\dv{\varphi_{i}}{t}=\eta_{\varphi,i}.
\end{equation}
Here, $\nu_a$ is the target
amplitude of the signals 
and $\tau_a$ is the damping timescale,
while $\eta_{a,i}$ and $\eta_{\varphi,i}$ are white Gaussian noises
and will be defined in a dimensionless form in the subsection F.

%//////////////////////////////////////////////////////////////////////////////////////////////////////////////////////////%
\subsection{Rankine vortex street}

In this and the following two subsections, we derive the hydrodynamic forces and torques in Eqs.~(\ref{EoMF}),~(\ref{EoMN}).
First, we formulate the vortex flow field.
Instead of 
the complex potential for a complete fluid~\cite{Ramananarivo2016,Oza2019},
we use the Rankine vortex for a viscous fluid,
which has a finite core radius and no singularity in the velocity field~\cite{Giaiotti2006}.
It is not only numerically tractable, but also gives a good representation of 
the cross-sectional velocity profile of the vortex ring shed by 
fish~\cite{Drucker1999,Lauder2002,Akanyeti2017}.
%{Giaiotti2006 は大気への応用で適当でない．Rankine vortex は古典的なので引用不要．}
The velocity field of a Rankine vortex whose center is located at ($X_0,Y_0$) 
is denoted by $\bm{u}_R(X-X_0,Y-Y_0; \Gamma_R)$, where
\begin{align}
\label{uR}
\bm{u}_R(X,Y; \Gamma_R)=\left\{ \begin{array}{ll}
\frac{\Gamma_R}{2\pi}\frac{(-Y,X)}{r_R^2}&[\sqrt{X^2+Y^2}\leq r_R],
\vspace{2mm}
\\
\frac{\Gamma_R}{2\pi}\frac{(-Y,X)}{X^2+Y^2}&[\sqrt{X^2+Y^2}>r_R].
\end{array} \right.
\end{align}
The magnitude of the velocity increases lineary with the distance from the center
within the core radius $r_R$,
while the velocity outside the core is described by a potential flow.
It gives the vorticity 
\begin{align}
\label{crossuR}
\grad\times\bm{u}_R(X,Y; \Gamma_R)=\left\{ \begin{array}{ll}
\frac{\Gamma_R}{\pi r_R^2}\bm{e}_Z&[\sqrt{X^2+Y^2}\leq r_R],\\
0&[\sqrt{X^2+Y^2}>r_R],
\end{array} \right.
\end{align}
and the circulation 
$\int_{\mathrm{core}}\dd X\dd Y {\grad\times\bm{u}_R}=\Gamma_R$.
\vspace{2mm}
%{$\Gamma_{\rm R}$ は符号付き．大きさ $\Gamma$ と区別する．→単位渦 $\hat{\bm{u}_R}$×循環 }

The magnitude of the circulation $\Gamma=|\Gamma_R|$ is estimated as follows.
According to the experiments on a flapping airfoil~\cite{Schnipper2009,Agre2016},
the circulation is estimated as
$\Gamma\approx C_\Gamma\pi^2A^2f/2$
with the help of the formula for the vorticity in the boundary layer.
Here, $A$ is the amplitude and $f$ is the frequency of flapping,
and $C_\Gamma\gtrsim1$ is 
a reasonable estimate for the prefactor~\cite{Schnipper2009,Agre2016}.
%{不等号では上限が分からない．$C_\Gamma = {\cal O}(1)$ としてよい？}
In our model, 
the amplitude $A_i(t)$ and frequency $f_i(t)$ 
of the flapping plate are time-dependent 
due to the physiological noise and hydrodynamic interaction,
and are calculated by Hilbert transformation (see Appendix~\ref{AppendHilbert}).
% (see Appendix A).
% the transverse displacement of the plate tip $y_{c,i}(t)=l_c\sin\theta_i(t)$
% by Hilbert transformation (see Appendix~\ref{AppendHilbert}),
% and the frequency is defined as $f_i(t)=\dv*{\phi_i}{t}$.
However, 
% $A_i(t)$ and $\phi_i(t)$ can not be obtained sequentially during a simulation
% due to Fourier transformations for time series data of $y_{c,i}(t)$ obtained after a simulation.
% Thus, we need to replace $A_i(t)$ and $f_i(t)$ by constants.
we will find that their deviations from the reference amplitude $A_0$ and the active frequency $f_a$ 
are small.
%{$A_0\sim 0.1 L_b$ は魚種によらない一定値．$\Gamma$ は定数であることを強調}
%We will obtain $A_i(t)\approx A_0$
%by tuning the amplitude of active torque $\nu_a$.
%Moreover, $f_i(t)\approx f_a$ as shown the following simulation results in the next section.
Therefore we can estimate the circulation as 
\begin{equation}
\label{Gamma}
\Gamma=
\frac{\pi^2}{2} C_\Gamma A_0^2f_a.
\end{equation}

Next, we introduce the Rankine vortex street.
A vortex is shed from a plate tip at the instant
when the angular velocity $\omega_i$ changes its sign.
When it changes from positive to negative
(which corresponds to a plate fully swung to the right),
the circulation of the vortex is $\Gamma_R = \Gamma$,
and $\Gamma_R = -\Gamma$ in the opposite case 
(corresponding to a plate fully swung to the left).
Subsequently, the vortex is carried away by the background flow $U$~\cite{Li2020}
and its strength decays exponentially in time~\cite{Oza2019}
with the timescale $\tau_\Gamma$.
The superposition principle can be applied to the vortex flow field %$\bm{u}(X,Y,t)$,
because the core radius $r_R$ is sufficiently small compared to
the distance between adjacent vortices $U/f_a$ and
the transverse distance $d_\perp$ between a pair of swimmers. % (see Fig.~\ref{schimaticswimmer}(b)).
Thus, the vortex flow field 
% $\bm{u}(X,Y,t)$ 
is given by
\begin{eqnarray}
\label{vortexfield}
\bm{u}(X,Y,t)&=&\sum_{i=1,2}\sum_{n_i}\exp(-\frac{t-t_{n_i}}{\tau_\Gamma})\nonumber\\
&& \quad \times\bm{u}_R(X-X_{n_i}(t),Y-Y_{n_i};s_{n_i}\Gamma).
\end{eqnarray}
Here $n_i$ is the index of the vortex shed by the swimmer $i$,
which is shed at the time $t_{n_i}$ and has the sign of circulation $s_{n_i}=\pm1$.
The position of the center of the vortex is given by %$n_i$ at time $t$ is
\begin{equation}
X_{n_i}(t)=X_i(t_{n_i})+\frac{L_b}{2}-l_c(1-\cos\theta_i(t_{n_i}))+U(t-t_{n_i}),
\end{equation}
\begin{equation}
Y_{n_i}=Y_i(t_{n_i})+l_c\sin\theta_i(t_{n_i}).
\end{equation}

%//////////////////////////////////////////////////////////////////////////////////////////////////////////////////////////%
\subsection{Drag and lift forces}

\begin{figure}[!t]
\centering
\includegraphics[width=\linewidth]{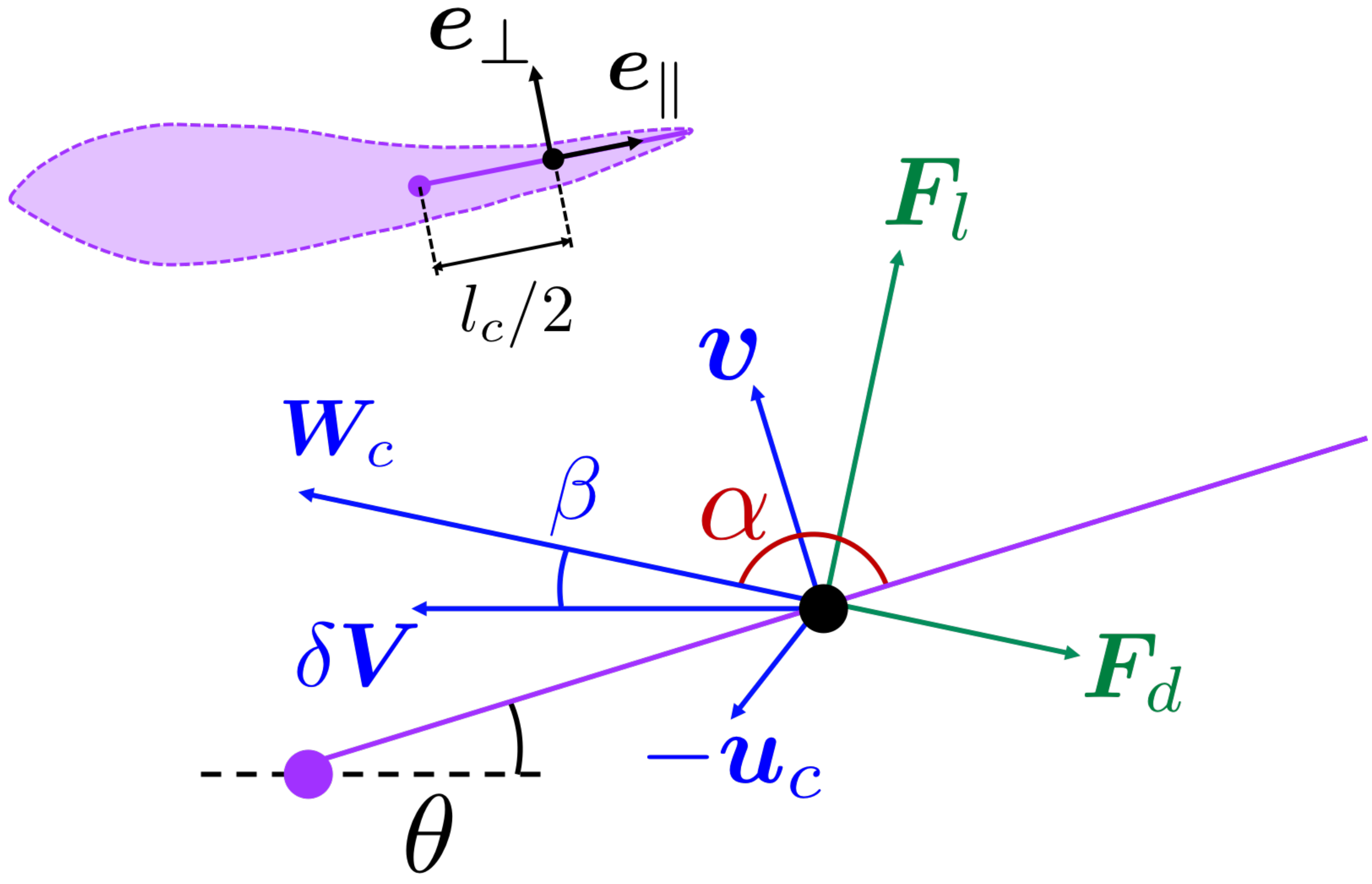}
%bb=0 0 360 252
\caption{
Schematic illustration of the flapping plate showing 
the relative velocity $\bm{W}_c$ and its relation to the drag and lift forces $\bm{F}_d,~\bm{F}_l$.
}
\label{dragliftfig}
\end{figure}

Next we calculate the drag and lift forces acting on the flapping plate.
In the quasi-steady approximation,
we estimate
these forces by Newton's drag law 
%at high Reynolds number%~\cite{Landau1987}
assuming that the plate flaps in the steady background
flow~\cite{Taylor1952}.
The relative velocity between a
plate and fluid is calculated at the center of pressure.
%of a two dimensional airfoil,
If the plate is infinitely high, the center of pressure
is located at the distance $l_c/4$ ($l_c$: plate length)
from the leading edge
as derived from the Joukowski theorem~\cite{Landau1987}.
In our case, the plate has a small finite aspect ratio ($H_b/l_c\lesssim1$),
and the center of pressure is approximately located at the center of the plate~\cite{Ortiz2015},
which has the coordinates 
\begin{equation}
X_{c,i}=X_i+\frac{L_b}{2}-l_c\leri{1-\frac{1}{2}\cos\theta_i},~Y_{c,i}=Y_i+\frac{l_c}{2}\sin\theta_i.
\end{equation}
Therefore, as shown in Fig.~\ref{dragliftfig}, we define the relative velocity as
\begin{equation}
\bm{W}_{c,i}=\bm{v}_i+\delta\bm{V}_i-\bm{u}_{c,i},
\end{equation}
where $\bm{v}_i$ is the rotational velocity of the center of the plate
\begin{equation}
\bm{v}_i=v_i\bm{e}_{\perp,i}, v_i=\frac{l_c}{2}\omega_i,
\end{equation}
with $\bm{e}_{\perp,i}=(-\sin\theta_i,\cos\theta_i)$ being the unit vector perpendicular to the plate,
$\delta\bm{V}_i=\bm{V}_i-\bm{U}=(V_i-U)\bm{e}_X$
is the thrust velocity of the swimmer, and
$\bm{u}_{c,i}=\bm{u}(X_{c,i},Y_{c,i},t)$
is the vortex flow velocity at the center of the plate.

Now we define the angle $\beta_i\in[-\pi,\pi]$
between $\bm{W}_{c,i}$ and $-\bm{e}_X$ (see Fig.~\ref{dragliftfig}).
By definition, $\beta_i$ is positive when $\bm{W}_{c,i}\cdot\bm{e}_Y>0$.
%Furthermore, 
It gives the angle of attack $\alpha_i\in[0,\pi]$ of the plate %for $\bm{W}_{c,i}$ 
as
\begin{equation}
\alpha_i=\pi\left\lceil\frac{\theta_i+\beta_i}{\pi}\right\rceil-(\theta_i+\beta_i),
\end{equation}
where $\lceil\circ\rceil$ is the ceiling function.
For example, in the case of Fig.~\ref{dragliftfig}, 
we have $\alpha=\pi-(\theta+\beta)$ because $0<\theta+\beta<\pi$.
%Based on the above,
Using these,
we can express the drag force $\bm{F}_{d,i}$ and the lift force $\bm{F}_{l,i}$
by Newton's drag law, as
\begin{equation}
\label{Fd}
\bm{F}_{d,i}=\frac{\rho_w}{2}C_d(\alpha_i)H_bl_cW_{c,i}^2\bm{e}_{d,i},
\end{equation}
\begin{equation}
\label{Fl}
\bm{F}_{l,i}=\frac{\rho_w}{2}C_l(\alpha_i)H_bl_cW_{c,i}^2\bm{e}_{l,i},
\end{equation}
\begin{equation}
\bm{e}_{d,i}=(\cos\beta_i,-\sin\beta_i),~\bm{e}_{l,i}=(\sin\beta_i,\cos\beta_i).
\end{equation}
Here,  the unit vectors $\bm{e}_{d,i}$ and $\bm{e}_{l,i}$ are parallel and perpendicular
to the relative velocity, respectively.
The drag coefficient $C_d(\alpha_i)$ and the lift coefficient $C_l(\alpha_i)$
are defined using previous results on airfoils; see Appendix~\ref{AppendCdClK} for details.

The thrust forces $F_{d,i}$ and $F_{l,i}$ in Eq.~(\ref{EoMF}) 
are the $X$-components of $\bm{F}_{d,i}$ and $\bm{F}_{l,i}$, 
and their sum reads
\begin{equation}
\label{FdFl}
F_{d,i}+F_{l,i}=(\bm{F}_{d,i}+\bm{F}_{l,i})\cdot\bm{e}_X=
\frac{\rho_w}{2}C_F(\alpha_i,\beta_i)H_bl_cW_{c,i}^2,
\end{equation}
\begin{equation}
C_F(\alpha_i,\beta_i)=C_d(\alpha_i)\cos\beta_i+C_l(\alpha_i)\sin\beta_i.
\end{equation}
The hydrodynamic torque $\bm{N}_{d,i}+\bm{N}_{l,i}$ in Eq.~(\ref{EoMN}) is given by
\begin{eqnarray}
\label{NdNl}
\bm{N}_{d,i}+\bm{N}_{l,i}&=&\frac{l_c}{2}\bm{e}_{\parallel,i}\times(\bm{F}_{d,i}+\bm{F}_{l,i})\nonumber\\
&=&\frac{\rho_w}{4}C_N(\alpha_i,\theta_i+\beta_i)H_bl_c^2W_{c,i}^2\bm{e}_Z,
\end{eqnarray}
\begin{eqnarray}
C_N(\alpha_i,\theta_i+\beta_i)&=&-C_d(\alpha_i)\sin(\theta_i+\beta_i)\nonumber\\
&&\quad +C_l(\alpha_i)\cos(\theta_i+\beta_i),
\end{eqnarray}
where $\bm{e}_{\parallel,i}=(\cos\theta_i,\sin\theta_i)$ is
the unit vector parallel to the plate (see Fig.~\ref{dragliftfig}).

The third force in Eq.~(\ref{EoMF}) 
is the Newton's drag force acting on the swimmer's body
\begin{equation}
\label{FD}
F_{D,i}=-\sgn{W_{b,i}}\rho_wC_DH_bL_bW_{b,i}^2,
\end{equation}
where $C_D$ is the drag coefficient,
$\sgn{\circ}$ is the sign function
\begin{align}
\label{sgnfunc}
\sgn{x}=\left\{ \begin{array}{ll}
x/\abs{x}&[x\neq0],\\
0&[x=0],
\end{array} \right.
\end{align}
and
\begin{equation}
W_{b,i}=\delta V_i-\bm{e}_X\cdot\bm{u}(X_i,Y_i,t)
\end{equation}
is the relative velocity of the swimmer  at the center of the body.

%//////////////////////////////////////////////////////////////////////////////////////////////////////////////////////////%
\subsection{Inertial force and added mass}

In the quasi-steady approximation, we add the inertial force
as a non-steady term in the equation of motion~\cite{Lighthill1960}.
Since the inertial force is proportional to the acceleration,
it is incorporated as an additional effective mass of the plate which is called the added mass~\cite{Landau1987}.
The added mass of an oscillating plate of finite aspect ratio is given by 
\begin{equation}
\label{addedmass}
m_p=\frac{\pi}{4}K\rho_wH_b^2l_c,
\end{equation}
where the numerical prefactor $K$ depends on the aspect ratio $H_b/l_c$
and is determined by fitting experimental data on airfoils~\cite{Brennen1982};
see Appendix~\ref{AppendCdClK} for details.

For the acceleration of the plate, we use the time-derivative of 
the velocity component that is perpendicular to the plate
$(\bm{v}_i+\delta\bm{V}_i)\cdot\bm{e}_{\perp,i}$.
Then we obtain the inertial force as  
\begin{equation}
\bm{F}_{m,i}=-m_p\dv{(v_i-\delta V_i\sin\theta_i)}{t}\bm{e}_{\perp,i},
\end{equation}
which gives the thrust force $F_{m,i}$ in Eq.~(\ref{EoMF}) as
\begin{equation}
\label{Fm}
F_{m,i}=\bm{F}_{m,i}\cdot\bm{e}_X=m_p\sin\theta_i\dv{(v_i-\delta V_i\sin\theta_i)}{t},
\end{equation}
and the torque $\bm{N}_{m,i}$ in Eq.~(\ref{EoMN}) as
\begin{equation}
\label{Nm}
\bm{N}_{m,i}=\frac{l_c}{2}\bm{e}_{\parallel,i}\times\bm{F}_{m,i}
=-\frac{m_p}{2}l_c\dv{(v_i-\delta V_i\sin\theta_i)}{t}\bm{e}_Z.
\end{equation}

%//////////////////////////////////////////////////////////////////////////////////////////////////////////////////////////%
\subsection{Non-dimensionalization of equations}

%Based on the above, we organize the equations.
We reorganize the equations in a non-dimensional form.
% First of all,  We define the length, time, and mass unit for non-dimensionalization:
The unit of length is the body length $L_b$,
the unit of time is taken as $\tau_0=1$ sec,
and the unit mass unit is the body mass $M$.
In the following, except for Appendix~\ref{AppendExperi},
all quantities are non-dimensionalized by $L_b$, $\tau_0$, and $M$
unless otherwise stated, and expressed by the same symbols as before.
(For example, we reexpress the dimensionless thrust speed $\delta V_i\tau_0/L_b$
by $\delta V_i$ 
and the bending stiffness $B\tau_0^2/ML_b^3$ by $B$.)
In addition, we define the dimensionless constants % for clarity:
\begin{equation}
\chi_h= \frac{H_b}{L_b}, \quad \chi_c= \frac{l_c}{L_b}, \quad \chi_\rho=\frac{\rho_w}{\rho}.
\end{equation}

Using the thrust forces (\ref{FdFl}),~(\ref{FD}) and~(\ref{Fm}),
and the torques (\ref{Ne}),~(\ref{Na}),~(\ref{NdNl}) and~(\ref{Nm}),
Eqs.~(\ref{EoMF}) and~(\ref{EoMN}) are rewritten as
\begin{equation}
\label{EoM1}
\mathcal{M}_i\dv{\delta V_i}{t}=\widehat{\mathcal{I}}_i\dv{\omega_i}{t}+\mathcal{F}_i,
\end{equation}
\begin{equation}
\label{EoM2}
\mathcal{I}\dv{\omega_i}{t}=\widehat{\mathcal{M}}_i\dv{\delta V_i}{t}+\mathcal{N}_i,
\end{equation}
respectively, where
\begin{widetext}
\begin{equation}
\mathcal{M}_i=1+\frac{\pi}{4}\chi_\rho\chi_c\chi_hK\sin^2\theta_i
,~\mathcal{I}=1+\frac{3\pi}{16}\chi_\rho\chi_hK
,~\widehat{\mathcal{M}}_i=\frac{3\pi}{8}\frac{\chi_\rho\chi_h}{\chi_c}K\sin\theta_i
,~\widehat{\mathcal{I}}_i=\frac{\pi}{8}\chi_\rho\chi_c^2\chi_hK\sin\theta_i,
\end{equation}
\begin{equation}
\label{Fi}
\mathcal{F}_i=\frac{1}{2}\chi_\rho\chi_cC_F(\alpha_i,\beta_i)W_{c,i}^2
-\frac{\pi}{4}\chi_\rho\chi_c\chi_hK\omega_i\delta V_i\sin\theta_i\cos\theta_i
-\sgn{W_{b,i}}\chi_\rho C_DW_{b,i}^2,
\end{equation}
\begin{equation}
\label{Ni}
\mathcal{N}_i=\frac{3}{4}\frac{\chi_\rho}{\chi_c}C_N(\alpha_i,\theta_i+\beta_i)W_{c,i}^2
+\frac{3\pi}{8}\frac{\chi_\rho\chi_h}{\chi_c}K\omega_i\delta V_i\cos\theta_i
-3\frac{B}{\chi_c^4}\sin\theta_i
+3\frac{N_{a,i}}{\chi_c^3}\sin(2\pi f_at+\varphi_i).
\end{equation}
\end{widetext}
We rewrite Eqs.~(\ref{EoM1}) and~(\ref{EoM2})
in the matrix form %as a matrix formation to make it numerially solvable:
\begin{equation}
\label{EoM}
\dv{t}\mqty[\delta V_i\\\omega_i]=
\frac{1}{\mathcal{M}_i\mathcal{I}-\widehat{\mathcal{M}}_i\widehat{\mathcal{I}}_i}
\mqty[\mathcal{I}&\widehat{\mathcal{I}}_i\\\widehat{\mathcal{M}}_i&\mathcal{M}_i]
\mqty[\mathcal{F}_i\\\mathcal{N}_i],
\end{equation}
where 
\begin{equation}
\mathcal{M}_i\mathcal{I}-\widehat{\mathcal{M}}_i\widehat{\mathcal{I}}_i=
1+\frac{\pi}{4}\chi_\rho\chi_hK\leri{\frac{3}{4}+\chi_c\sin^2\theta_i}>0.
\end{equation}
The stochastic differential equations (\ref{Nanoise}) and (\ref{phinoise}) 
are rewritten as
\begin{equation}
\label{stNa}
\dd N_{a,i}=\frac{\dd t}{\tau_a}(\nu_a-N_{a,i})+\sqrt{2D_a}\dd w_{a,i},
\end{equation}
\begin{equation}
\label{stphi}
\dd\varphi_i=\sqrt{2D_\varphi}\dd w_{\varphi,i},
\end{equation}
where $\dd w_{a,i}$ and $\dd w_{\varphi,i}$ are the standard Wiener processes,
and $D_a$ and $D_\varphi$ are the diffusion coefficients.

%//////////////////////////////////////////////////////////////////////////////////////////////////////////////////////////%
\subsection{Numerical method}%Settings of the simulation}

% From the above,
We numerically solve Eqs.~(\ref{EoM}), (\ref{stNa}), and~(\ref{stphi})
with the vortex flow field (\ref{vortexfield}).
We integrate the equation of motion (\ref{EoM}) by the Euler method 
with the time step $\Delta t=0.0005$.
For the stochastic differential equations,
we use It$\hat{\rm{o}}$ integral % method which solves 
%with additive noise. % ~\cite{McKean1969}. 
and replace the standard Wiener process by $\xi \sqrt{\Delta t}$, %\times\mathrm{N}(0,1)$,
where $\xi$ %$\mathrm{N}(0,1)$ 
is a random number generated by the standard Gauss distribution
that is truncated at $\pm 5 \sigma$ to prevent divergence of the solution.
%To prevent divergence of the solution by %anomaly
%outliers in the noises,
%we replace the random numbers in $\mathrm{N}(0,1)$ 
%that are greater than $5$ or less than $-5$ by $\pm5$, 
%respectively. 
%Note that such an event occurs only at the rate of $6\times10^{-7}$\%, and 
%therefore it is not practically issue.
%does not affect the results in any significant way.
We also introduce a finite lifetime for the vortices
to reduce computational cost. The vortex $n_i$ is deleted
when it satisfies the condition $\exp(-(t-t_{n_i})/\tau_\Gamma)\leq10^{-3}$.

For the initial conditions,
%the common conditions for fish1 and fish2 are
we used
$\delta V_i=-U$, $\theta_i=0$, $\omega_i=0$, and $N_{a,i}=\nu_a$ 
for both fish1 ($i=1$) and fish2 ($i=2$),
while $\varphi_i$ is 
chosen as a uniform random number in $[0,2\pi]$.
Fish1 has the initial position $(X_1,Y_1)=(0,0)$,
while for fish2, $X_2$ is chosen as a uniform random number
in $[-d_{\parallel,\max},d_{\parallel,\max}]$ with $d_{\parallel,\max}=2.5$
and $Y_2=d_\perp$ (see Fig.~\ref{schimaticswimmer}(b)).

Each simulation runs up to $t_{\max}=80$.
We obtain the amplitude $A_i(t)$ and the phase $\phi_i(t)$ of the flapping plate
by Hilbert transformation of the tip position $y_{c,i}(t)$ 
%by fast Fourier transformation
in the time window $[t_{\max}-t_H,t_{\max}]$
with $t_H=2^{17}\times\Delta t=65.536$ (see also Appendix~\ref{AppendHilbert}),
and computed the frequency %$f_i(t)$ by 1st order difference 
$f_i(t)=\{\phi_i(t)-\phi_i(t-\Delta t)\}/\Delta t$.
We confirmed that a swimmer rapidly reaches steady swimming by $t=t_{\max}-t_H$
in the noiseless case.
To avoid artifacts of the Hilbert transformation 
%is not going well numerically
at both ends of the time interval, %$t\in[t_{\max}-t_H,t_{\max}]$,
%Note that
% because of the premise that
%$y_{c,i}$ is a periodic function with a period $t_H$ in Fourier transformation,
%We, therefore, 
we introduce the cutoff $\delta t$
and use the interval $t\in[t_{\max}-t_H+\delta t,t_{\max}-\delta t]$
for time-averaging.
%. some quantities.
We denote the time-average of the quantity $Q(t)$ by $\overline{Q}$.
% in this time interval.
%(The value of $\delta t$ depends on the specific cases.)

In this model, the primary control parameters are
$\{\chi_c,~C_\Gamma,~f_a,~D_a,~D_\varphi,~d_\perp\}$,
and $U$ and $\nu_a$ are varied depending on the case. %sse parameter value.
We set the primary control parameters %The range of these parameters are based on corresponding 
in accordance with experimental data;
we summarize the parameters and their values used in the simulation in Appendix~\ref{AppendExperi}.

%%%%%%%%%%%%%%%%%%%%%%%%%%%%%%%%%%%%%%%%%%%%%%%%%%%%%%%%%%%%
\section{Solitary swimming}%Comparison with experiments for solitary swimming}

In this section, we show the results for solitary swimming.
We omit the index $i$ as we consider only one fish.

%Henceforward, we will show the results of the simulation.
%In this section, we omit the index $i=1,2$ of all quantities because
%we treat solitary fish.

\begin{figure}[!b]
\centering
\includegraphics[width=\linewidth]{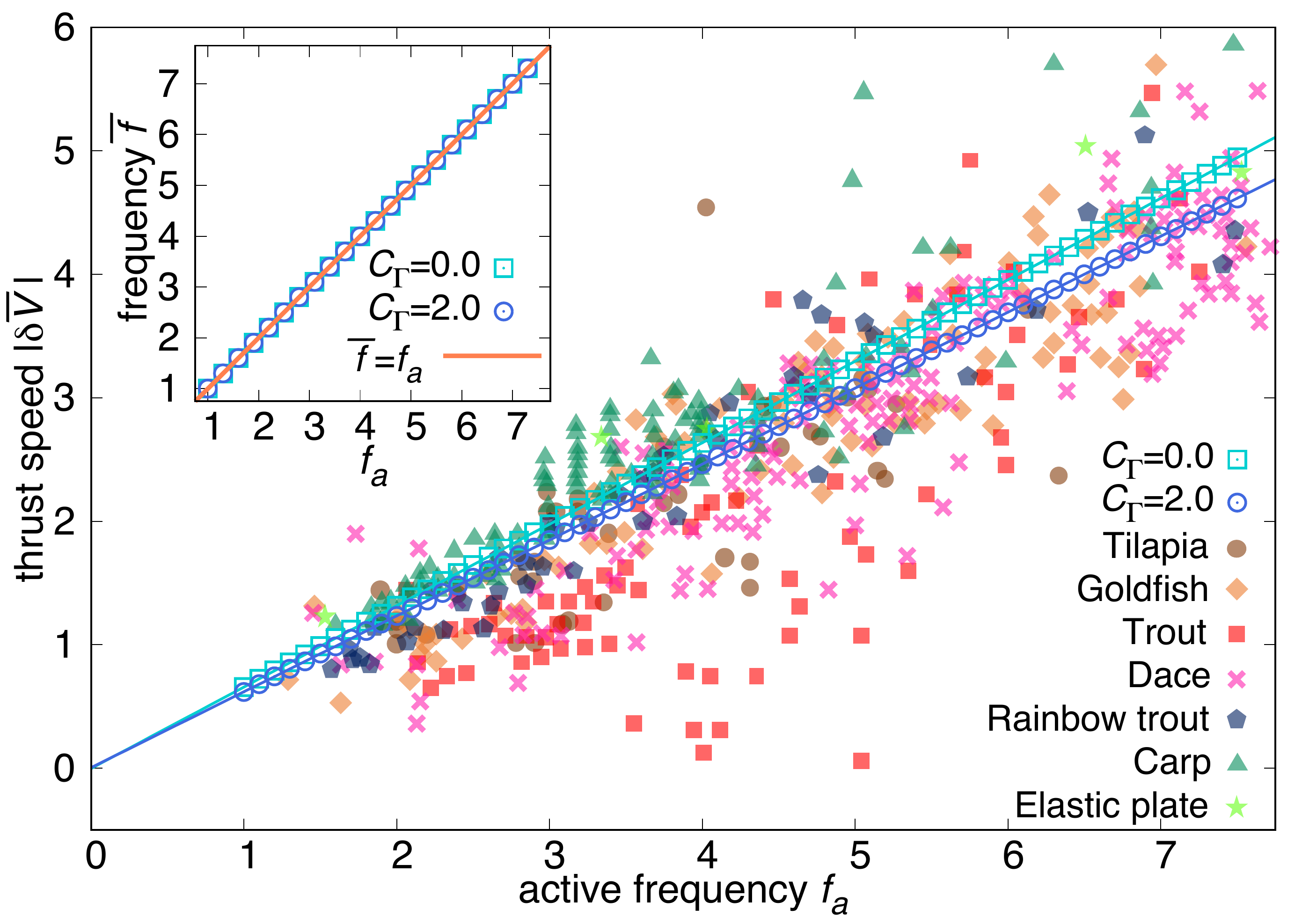}
%bb=0 0 360 252
\caption{
The thrust speed $\abs{\delta\overline{V}}$ as
a function of the active frequency $f_a$.
Open squares and open circles correspond to
the case without vortices ($C_\Gamma=0.0$) and with vortices ($C_\Gamma=2.0$), respectively,
and solid lines show the fitting by Eq.~(\ref{Vmufa}) (see also Table~\ref{mutable}).
The other types of points represent the experimental data 
for tilapia and carp~\cite{Tanaka1996},
goldfish, trout, and dace~\cite{Bainbridge1958},
rainbow trout~\cite{Akanyeti2017}, 
and the data for the elastic plate model~\cite{Gazzola2015}.
In the inset, we show
the time-averaged frequency $\overline{f}$ versus  $f_a$, 
which is perfectly on the line $\overline{f}=f_a$.
}
\label{Vffa}
\end{figure}

%//////////////////////////////////////////////////////////////////////////////////////////////////////////////////////////%
\subsection{The relation between thrust speed and tailbeat frequency}

First we consider the noiseless case ($D_a=0,~D_\varphi=0$),
for which the parameter set is $\{\chi_c,~C_\Gamma,~f_a\}$.
We set $U=0$ by a Galilean transformation and without loss of generality, so that $V=\delta V$.
For each parameter set $\{\chi_c,~C_\Gamma,~f_a\}$, 
we tune $\nu_a$ in increments of $\Delta\nu_a=0.01$
so that the time-averaged amplitude $\overline{A}$ becomes
close to the prescribed value $A_0=0.1L_b$.
%{$A_0/L_b \sim 0.1$ は魚種によらず一定であることを再現するため，$\nu_A$ を調節する}

To test the validity of the model,
we study the relation between the time-averaged thrust speed $\abs{\delta\overline{V}}$
and the active frequency $f_a=1.0$-7.5, with $\chi_c=0.375$ fixed.
%(We select the cutoff time $\delta t=1/f_{a,\min}=1.0$.)
%{cuttoff time 言及不要}

As shown in Fig.~\ref{Vffa},
$\abs{\delta\overline{V}}$ increases linearly as a function of $f_a$,
for the cases without vortices ($C_\Gamma=0.0$) and
with vortices ($C_\Gamma=2.0$).
The time-averaged tailbeat frequency $\overline{f}$
derived from Hilbert transformation is almost equal to the active frequency $f_a$
(see Fig.~\ref{Vffa} inset).
A linear relation between the thrust speed and tailbeat frequency 
was found in many experimental studies~\cite{Bainbridge1958,Hunter1971,Webb1984,
Akanyeti2017,Li2021,Nagai1979,Tanaka1996}.
Our data are also nicely fitted
by the linear relation
\begin{equation}
\label{Vmufa}
\abs{\delta\overline{V}}=\mu f_a+\mu',
\end{equation}
where $\mu$ and $\mu'$ are constants.
In Table~\ref{mutable}, we show our results in comparison with
the experimental data~\cite{Tanaka1996,Nagai1979,Bainbridge1958,Akanyeti2017}
and the numerical results for the elastic plate model~\cite{Gazzola2015}.
For both with and without vortices, the values of $\mu$ and $\mu'$ 
fit in the range of the previous results.
In particular, $\mu$ lies between 0.6 and 0.7 
regardless of the presence of vortices,
and is close to the experimental values.
The intercept $\mu'$ is almost zero (slightly negative).
This is in agreement with the experimental results
for five species, where $\mu'=0$ is assumed~\cite{Tanaka1996,Nagai1979}.
On the other hand, two other experiments obtained 
negative values of $\mu'$~\cite{Bainbridge1958,Akanyeti2017}.
The origin of the negative intercept is unclear, 
but we may argue that
the non-caudal fins (e.g. the pectoral fin) that rise from 
the body in low speed swimming~\cite{Bainbridge1963}
induce additional drag forces and dampen the thrust speed 
to zero at a finite tailbeat frequency.

\begin{table}[!t]
\begin{center}
\caption{
 The coefficients $\mu$ and $\mu'$ in Eq.~(\ref{Vmufa}).
 Our results are shown in the top two rows.
 The rest rows are results of experiments and other model.
The data (T), (N), (B), and (A) are taken from the experiments
 in Ref.~\cite{Tanaka1996},~\cite{Nagai1979},~\cite{Bainbridge1958}, and~\cite{Akanyeti2017},
 respectively, and (G) is from the elastic plate model~\cite{Gazzola2015}.
}
\begin{tabular}{wc{40mm}|wc{20mm}wc{20mm}}
& $\mu$ & $\mu'$\\\hline\hline
without vortices ($C_\Gamma=0.0$) & 0.66 & $-0.002$\\
with vortices ($C_\Gamma=2.0$) & 0.61 & $-0.004$\\\hline
Tilapia (T)&0.576&0.0\\
Goldfish (N)&0.61&0.0\\
Goldfish (B)&0.64&$-0.20$\\
Trout (N)&0.62&0.0\\
Trout (B)&0.73&$-1.13$\\
Dace (N)&0.63&0.0\\
Dace (B)&0.74&$-1.02$\\
Rainbow trout (A)&0.67&$-0.16$\\
Carp (T)&0.695&0.0\\
Elastic plate (G)&0.72&$-0.12$\\\hline
\end{tabular}
\label{mutable}
\end{center}
\end{table}

%//////////////////////////////////////////////////////////////////////////////////////////////////////////////////////////%
\subsection{Other properties of noiseless swimming}

%%%%%%%%%%%%%%%%%%%%%%%%%%%%%%%%%%%%%%%%%%%%%
\begin{figure}[!t]
\centering
\includegraphics[width=\linewidth]{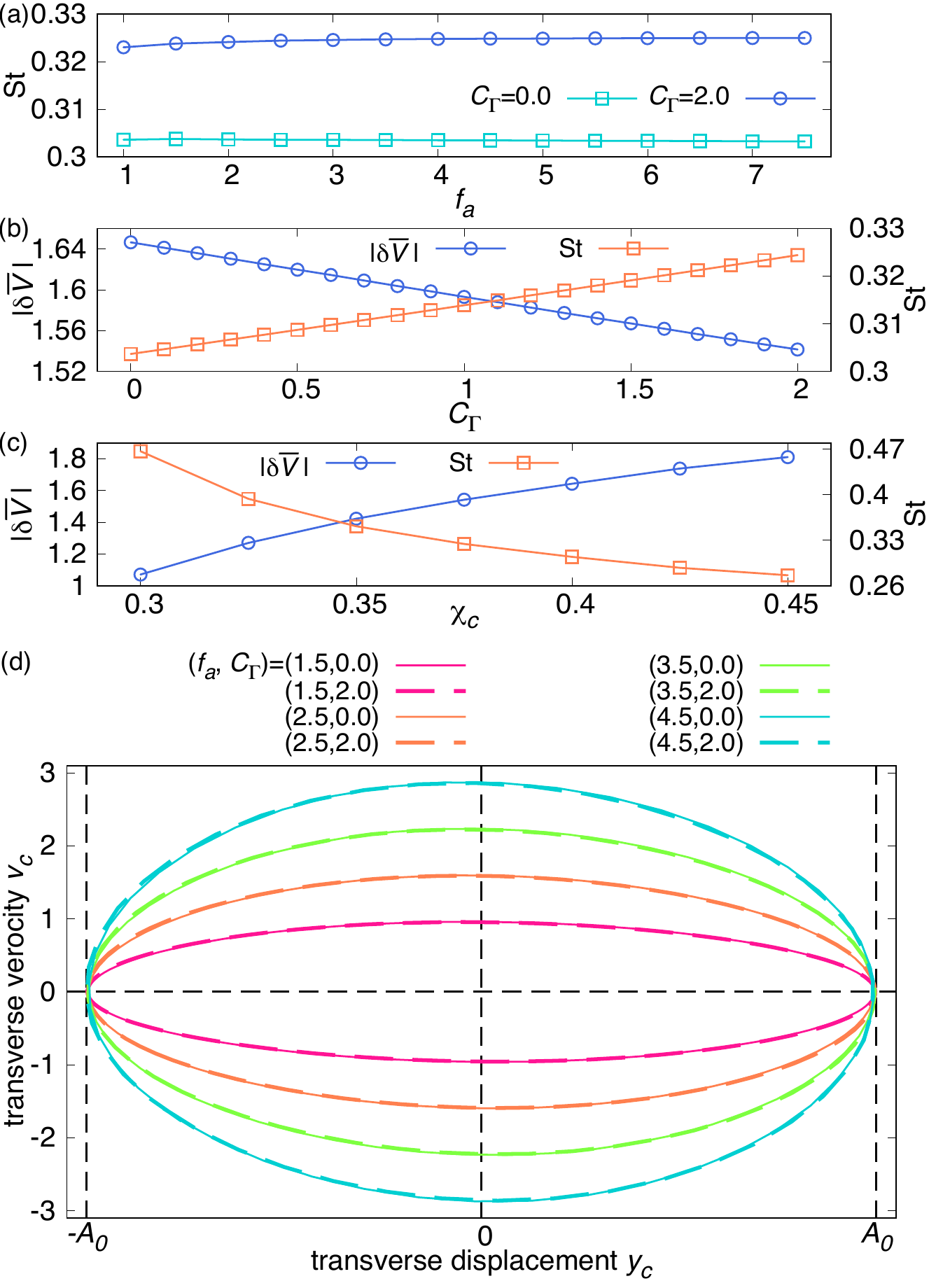}
%bb=0 0 360 252
\caption{
(a)-(c) The Strouhal number $\HySt$ and thrust speed $\abs{\delta\overline{V}}$ as functions of $f_a$, $C_\Gamma$, and $\chi_c$.
(a)  Square and circle points correspond to
the case without vortices ($C_\Gamma=0.0$) and with vortices ($C_\Gamma=2.0$), respectively.
We fixed $\chi_c=0.375$.
(b)-(c) Circle points represent $\abs{\delta\overline{V}}$ and 
square points correspond to $\HySt$.
In (b), we fixed $f_a=2.5$ and $\chi_c=0.375$;
in (c), $f_a=2.5$ and $C_\Gamma=2.0$.
(d) The trajectory of tailbeat in phase space ($y_c,v_c$) 
for $\chi_c=0.375$.
Solid and dashed lines correspond to
the case without vortices ($C_\Gamma=0.0$) and with vortices ($C_\Gamma=2.0$), respectively.
A larger $f_a$ gives a larger amplitude of the velocity.
}
\label{StVorb}
\end{figure}
%%%%%%%%%%%%%%%%%%%%%%%%%%%%%%%%%%%%%%%%%%%%%

Let us study some more properties of noiseless swimming.
We introduce the Strouhal number
\begin{equation}
\label{Stdef}
\HySt=\frac{2\overline{A}\,\overline{f}}{\abs{\delta\overline{V}}},
\end{equation}
which characterizes the speed of flapping compared to the thrust.
(The amplitude $\overline{A}$ is doubled following convention.)
As shown in Fig.~\ref{StVorb}(a),
the Strouhal number is given by $\HySt\gtrsim0.3$
independent of $f_a$ and for $\chi_c=0.375$.
This value is agreement with the experimental values of $\HySt \approx0.2$-0.4,
which are often close to $0.3$
for many species~\cite{Gazzola2015,Gazzola2014,Triantafyllou1993,Taylor2003}.
%St = 0.3 Gazzola2014 Fig.2b
%St = 0.2-0.4 Gazzola2015 Sup 

Fig.~\ref{StVorb}(b) shows the dependence of $\abs{\delta\overline{V}}$ and $\HySt$
on the vortex strength $C_\Gamma$ with $f_a=2.5$ and $\chi_c=0.375$.
The thrust speed $\abs{\delta\overline{V}}$ gradually decreases as $C_\Gamma$ 
increases,
which is because the vortex flow increases the drag on the body and the plate.
As a result, the Strouhal number increases according to the definition (Eq.~(\ref{Stdef})).
%{式による細かい説明は分かりにくいので削除した．}
%%through the terms $\bm{W}_c$ and $W_b$ of Eqs.~(\ref{Fi}) and~(\ref{Ni})
%% that contains $\bm{u}\propto C_\Gamma$ as $\delta\bm{V} - \bm{u}$. 
%the vortex flow $\bm{u}\propto C_\Gamma$ is included
%always together with the thrust velocity $\delta\bm{V}$,
%and therefore
% $\abs{\delta\overline{V}}$ decreases linearly
% with increasing the drag on a body and a plate by the vortex flow.
In addition, we check the dependence of $\abs{\delta\overline{V}}$ and $\HySt$
on the relative fin length $\chi_c$; 
see Fig.~\ref{StVorb}(c).
As the caudal part becomes longer, 
the thrust force and speed $\abs{\delta\overline{V}}$ increase nonlinearly
due to the prefactor $K$ in 
the added mass (see Eq.~(\ref{addedmass}) and Fig.~\ref{CdClKfig}(b)).
The Strouhal number then decreases but stays in the 
experimentally observed range
$\HySt\approx0.2 - 0.4$~\cite{Gazzola2015,Gazzola2014,Triantafyllou1993,Taylor2003},
except for $\chi_c=0.3$.

Fig.~\ref{StVorb}(d) shows the tailbeat trajectory on the phase space $(y_c,v_c)$ 
%with $\chi_c=0.375$,
where $v_c=\dv*{y_c}{t}=l_c\omega\cos\theta$ is the transverse velocity.
There is almost no difference between
the trajectories in the case of $C_\Gamma=0.0$ and $C_\Gamma=2.0$.
The trajectory is almost mirror symmetric 
with respect to the $y_c$-axis and $v_c$-axis.
% with the amplitude $A_0$.
This symmetry of the caudal fin movement
is observed for 
steady swimming of fish~\cite{Bainbridge1963}.
Furthermore, the peak value of $\abs{v_c}$ is good agreement with
the experimental value: for example,
the peak value is $\abs{v_c}\sim1.5$-3.0 BL/s of dace with
the swimming speed $\gtrsim1.5$ BL/s~\cite{Bainbridge1963}.
It corresponds to the value for $f_a=2.5$-3.5 in Fig.~\ref{StVorb}(d)
(see also Fig.~\ref{Vffa} for the swimming speed in the range $f_a=2.5$-3.5).

%//////////////////////////////////////////////////////////////////////////////////////////////////////////////////////////%
\subsection{The effect of physiological noises}

%%%%%%%%%%%%%%%%%%%%%%%%%%%%%%%%%%%%%%%%%
\begin{figure}[!b]
\centering
\includegraphics[width=\linewidth]{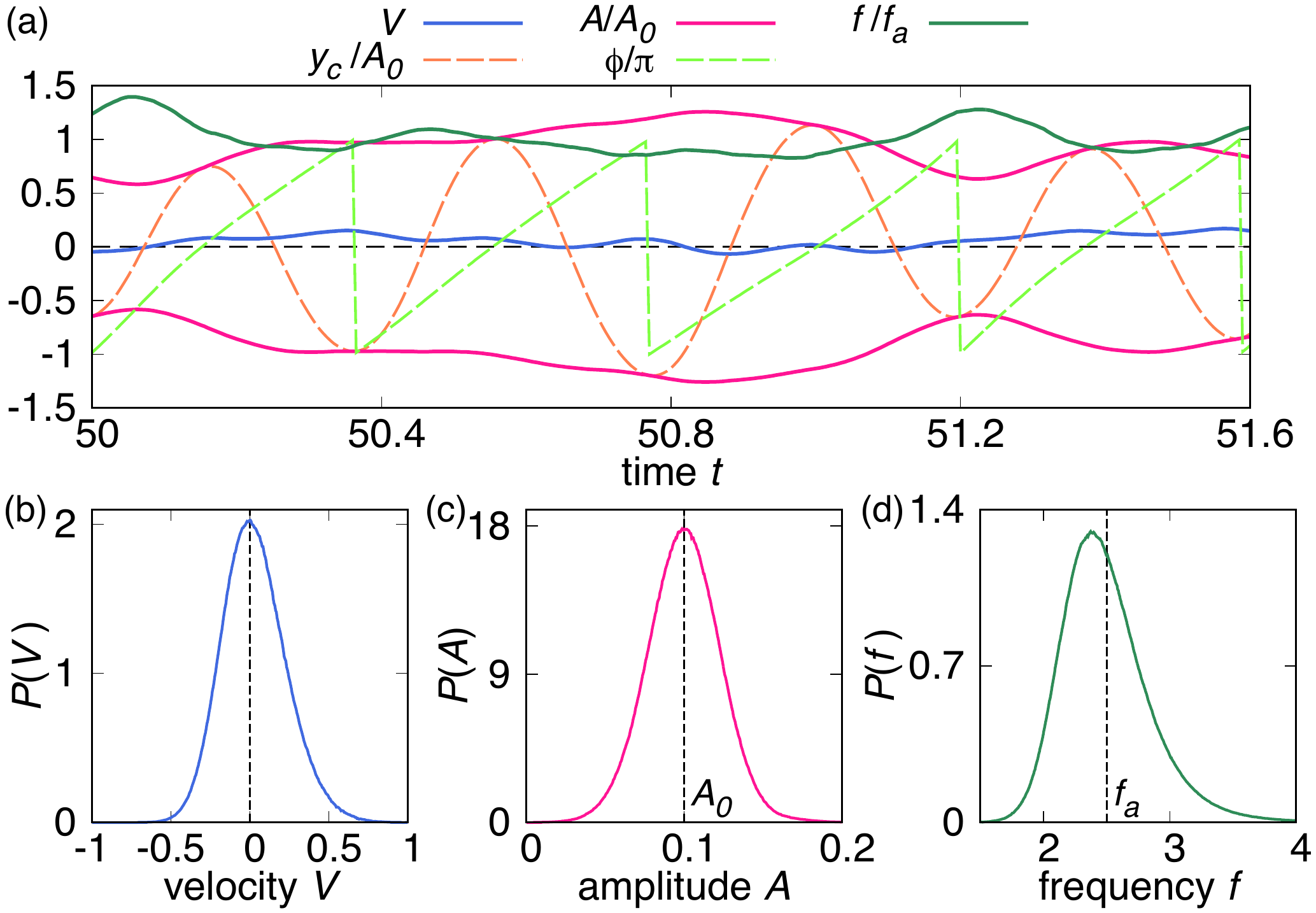}%eps}
%bb=0 0 360 252
\caption{
The effect of noise on swimming properties
with $C_\Gamma=2.0,D_a=0.7,D_\varphi=0.25$.
(a) Time evolution %of some quantities 
in the interval $4/f_a$.
The solid line near the dashed line of zero is $V$ and
The pair of solid lines are $A/A_0$ and $-A/A_0$.
The upper solitary solid line is $f/f_a$.
The sinusoidal dashed line is $y_c/A_0$ and
the sawtooth-like dashed line is $\phi/\pi$.
The normalized distribution $P$ of (b) the velocity $V$,
(c) the amplitude $A$ with dashed line $A_0$,
and (d) the frequency $f$ with dashed line $f_a$. 
}
\label{tevdis}
\end{figure}
%%%%%%%%%%%%%%%%%%%%%%%%%%%%%%%%%%%%%%%%%

Here, we study the effect of physiological noises. 
To make the expected value of $V$ close to zero,
we select the background flow speed $U=\abs{\delta V_0}$ 
where $\abs{\delta V_0}$ is the time averaged thrust speed $\abs{\delta\overline{V}}$
for the noiseless case.
Hereafter, we fix $\chi_c=0.375$, which reproduces
the thrust speed and active frequency,
and $f_a=2.5$, 
which corresponds to 
the background flow speed $U \sim 1.5-1.6$ BL/s in the experiment~\cite{Li2020};
see Fig.~\ref{Vffa}.
%(We use the time interval cut off $\delta t=1/f_a=0.4$.)
%$\chi_c=0.375$,  $f_a=2.5$, $D_a=0$, and $D_\varphi=0$.
%{\chi_c$ と $f_a$ の値は上で言った．$C_\Gamma$ は変えている．}

Fig.~\ref{tevdis}(a) shows the typical time evolution of
$V$ and normalized quantities $y_c/A_0,~A/A_0,~\phi/\pi,~f/f_a$
with noises.
We define the probability distribution of any quantity $Q$ in each run
as
\begin{equation}
\label{pQ}
p(Q)=\frac{1}{t_H-2\delta t}\int_{t_{\max}-t_H+\delta t}^{t_{\max}-\delta t}\dd t
\, \delta(Q-Q(t)),
\end{equation}
where $\delta(\circ)$ is Dirac delta function.
Then we take the ensemble average of $p(Q)$ over 1000 independent runs
to obtain the averaged probability distribution $P(Q)$.
Shown in Fig.~\ref{tevdis}(b)-(d) are the distributions
$P(V)$, $P(A)$, and $P(f)$ 
with the noise strengths $D_a=0.7$ and $D_\varphi=0.25$.
For these values of $D_a$ and $D_\varphi$,
the expected values of $A$ and $f$ are very close 
to the values $A_0$ and $f_a$ of the noiseless case,
and justify the estimate of $\Gamma$ in Eq.~(\ref{Gamma}).
Furthermore, the noise strengths nicely reproduce
the height, width, and asymmetry of the distributions observed for 
goldfish (Ref.\cite{Li2020}, Figs. 24 and 26 of the Supplementary Information).
Therefore, we choose $D_a=0.7,~D_\varphi=0.25$
as the standard parameter values in the simulations.
The dependence of the distributions on $D_a$ and $D_\varphi$
is shown in Appendix~\ref{Appenddistri}.

\begin{figure}[!t]
\centering
\includegraphics[width=\linewidth]{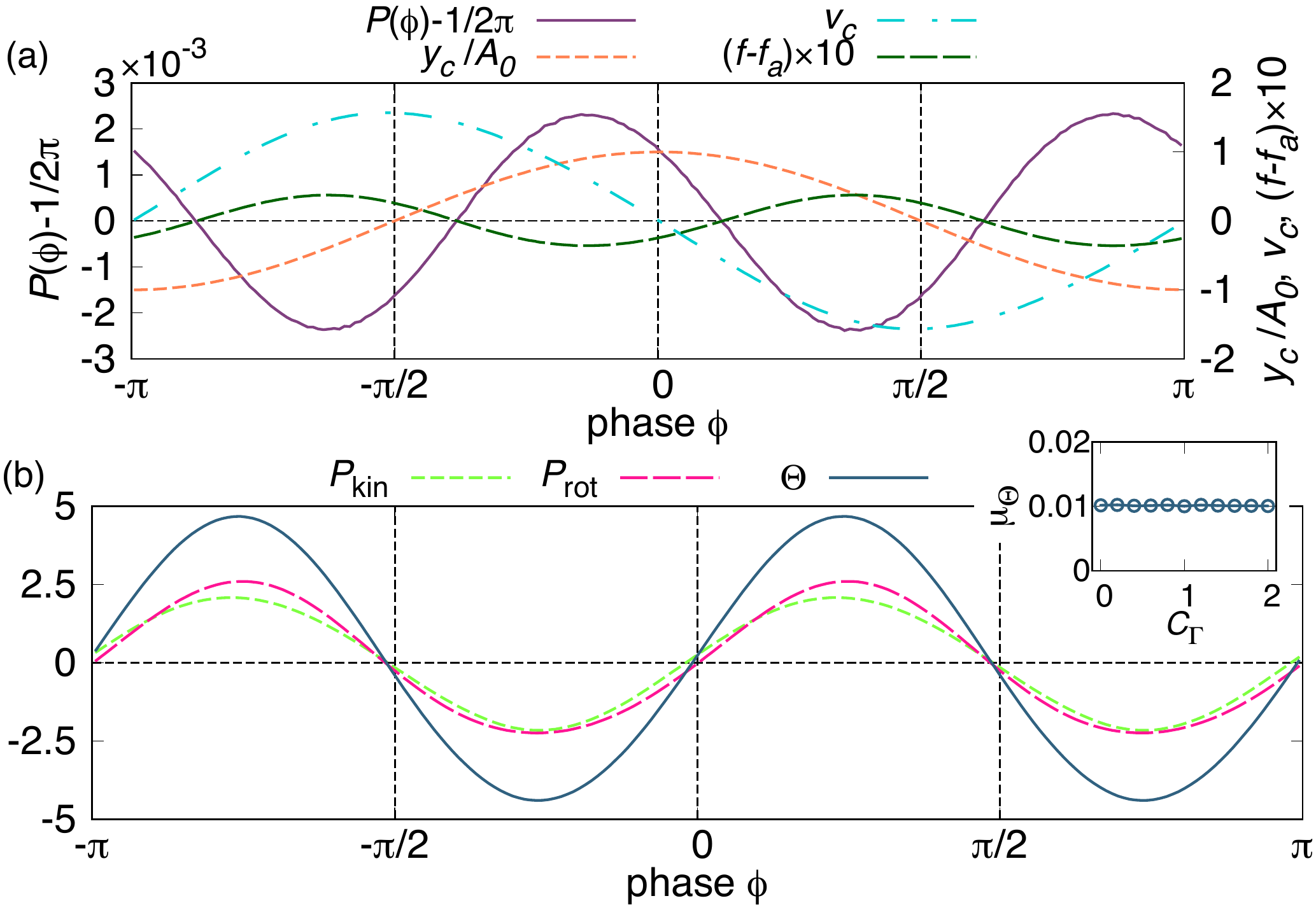}%eps}
%bb=0 0 360 252
\caption{
Profiles of various quantities characterizing the tailbeat 
as functions of the phase $\phi$, 
for $C_\Gamma=2.0$, $D_a=0.7$, and $D_\varphi=0.25$.
(a) The phase distribution $P(\phi)$ subtracted by its period average, 
transverse displacement $y_c/A_0$,
transverse velocity $v_c$, %(10 times of) 
frequency shift $f-f_a$ versus $\phi$.
(b) The power $P_{\mathrm{kin}}$ and $P_{\mathrm{rot}}$,
and the energy dissipation rate $\Theta$ versus $\phi$.
Inset: the dependence of $\mu_\Theta$ on $C_\Gamma$.
}
\label{Pphitail}
\end{figure}

Next, we consider the profiles of various quantities that characterize the tailbeat.  
%We define the distribution of the phase of tailbeat $P(\phi)$
%by averaging of $10^4$ simulations of $p(Q=\phi)$ (Eq.~(\ref{pQ})).
In Fig.~\ref{Pphitail}(a), we show the average phase distribution $P(\phi)$.
Its shift from the period average $1/(2\pi)$
slightly deviates from zero and oscillates with the period $\pi$.  
In addition, Fig.~\ref{Pphitail}(a) shows the transverse displacement $y_c/A_0$,
transverse velocity $v_c$, and frequency shift $f-f_a$ 
as functions of $\phi$:
these quantities are time-averages over the 
interval $t\in[t_{\max}-\delta t,t_{\max}-t_H+\delta t]$ and $10^4$ simulations
for each bin of $\phi$.
We confirm that $y_c/A_0$ is proportional to $\cos \phi$ 
by definition of the Hilbert transformation (see Appendix~\ref{AppendHilbert}),
and $v_c$ is proportional to $-\sin\phi$.
Roughly speaking, 
the phase distribution $P(\phi)$ is large when the plate is swinging away
from the midline of the body, and small when swinging back,
but there is a phase delay. In other words, the frequency (phase velocity) $f$
is small when the plate is swinging away, and vice versa.

To elucidate the reason of non-uniformity of $P(\phi)$,
we consider 
the energy dissipation rate $\Theta_{\mathrm{dis}}$,
or the power required for swimming.
It satisfies 
\begin{equation}
\Theta_{\mathrm{dis}} = \dv{E_{\mathrm{kin}}}{t}+\dv{E_{\mathrm{rot}}}{t},
\end{equation}
where 
$E_{\mathrm{kin}}=M(\delta V)^2/2$ is the translational kinetic energy of the body and
$E_{\mathrm{rot}}=I_c\omega^2/2$ is the rotational kinetic energy of the plate.
We divide the both sides of the equation by $ML_b^2/\tau_0^3$
and define 
the dimensionless energy dissipation rate
$\Theta= \Theta_{\mathrm{dis}}\tau_0^3/ML_b^2$, 
which satisfies
\begin{equation}
\label{Thetadef}
\Theta
=P_{\mathrm{kin}}+P_{\mathrm{rot}} \equiv
\delta V \dv{\delta V}{t} +
\frac{\chi_c^3}{3} \omega \dv{\omega}{t}.
\end{equation}
Note that $\Theta>0$ corresponds to the situation that a swimmer consumes the swimming energy. 
Fig.~\ref{Pphitail}(b) shows that $\Theta$, $P_{\mathrm{kin}}$, and $P_{\mathrm{rot}}$
oscillate with the period $\pi$ as a function of $\phi$.
We find that the energy dissipation rate
tends to be positive when $P(\phi)-1/2\pi<0$ (and $f-f_a>0$),
although there is a phase delay.
This result indicates that the swimmer consumes energy in 
quick motion of the caudal plate when it is swinging back to the midline,
and gain energy from the flow when the plate is swinging away.
The cycle-average of the energy dissipation rate,
defined by 
$\mu_\Theta = \int_{-\pi}^\pi d\phi \Theta(\phi) P(\phi)$,
is confirmed to be positive, 
but is very small compared to the amplitude of $\Theta$,
as shown in the inset of Fig.~\ref{Pphitail}(b).
We also find that $\mu_\Theta$ is almost independent of the vortex strength $C_\Gamma$.

%%%%%%%%%%%%%%%%%%%%%%%%%%%%%%%%%%%%%%%%%%%%%%%%%%%%%%%%%%%%
\section{Pair swimming}% and spontaneous optimization of energy}

In this section, we show the results for a pair of swimmers (labeled by $i=1,2$).
We fix $\chi_c=0.375$, $f_a=2.5$, $D_a=0.7$, and $D_\varphi=0.25$,
and choose the vortex strength
$C_\Gamma$ and the transverse 
distance between the swimmers $d_\perp$
as tunable control parameters.
We confirmed that our results do not change if fish 2 is positioned 
to the left of fish1, as expected from the left-right symmetry of the model 
(data not shown).

%//////////////////////////////////////////////////////////////////////////////////////////////////////////////////////////%
\subsection{Correllation between the phase difference and distance}

%We define the longitudinal distance  swimmers $d_\parallel=\abs{X_1-X_2}$
%(see Fig.~\ref{schimaticswimmer}(b)), and

First, we consider the phase difference 
between the phases of the tailbeat of the leader and follower,
defined by
\begin{align}
\psi=\left\{ 
\begin{array}{ll}
\phi_1-\phi_2&[X_1<X_2],\\
\phi_2-\phi_1&[X_2<X_1].
\end{array} \right.
\end{align}
%Note that the leader swimmer has smaller $X_i$ rather than the follower
%because the swimmers swim to $-\bm{e}_X$ direction.
To see the correlation between the phase difference and 
the longitudinal distance $d_\parallel=\abs{X_1-X_2}$,
we introduce
the conditional probability distribution $P(\psi; d_\parallel)$.
As a function of $\psi$, 
it is normalized in each bin of $d_\parallel$
by the condition $\int_{-\pi}^\pi\dd \psi P(\psi;d_\parallel)=1$.
Note that $d_\parallel$ and $\psi$ change with time in a simulation.

\begin{figure}[!t]
\centering
\includegraphics[width=\linewidth]{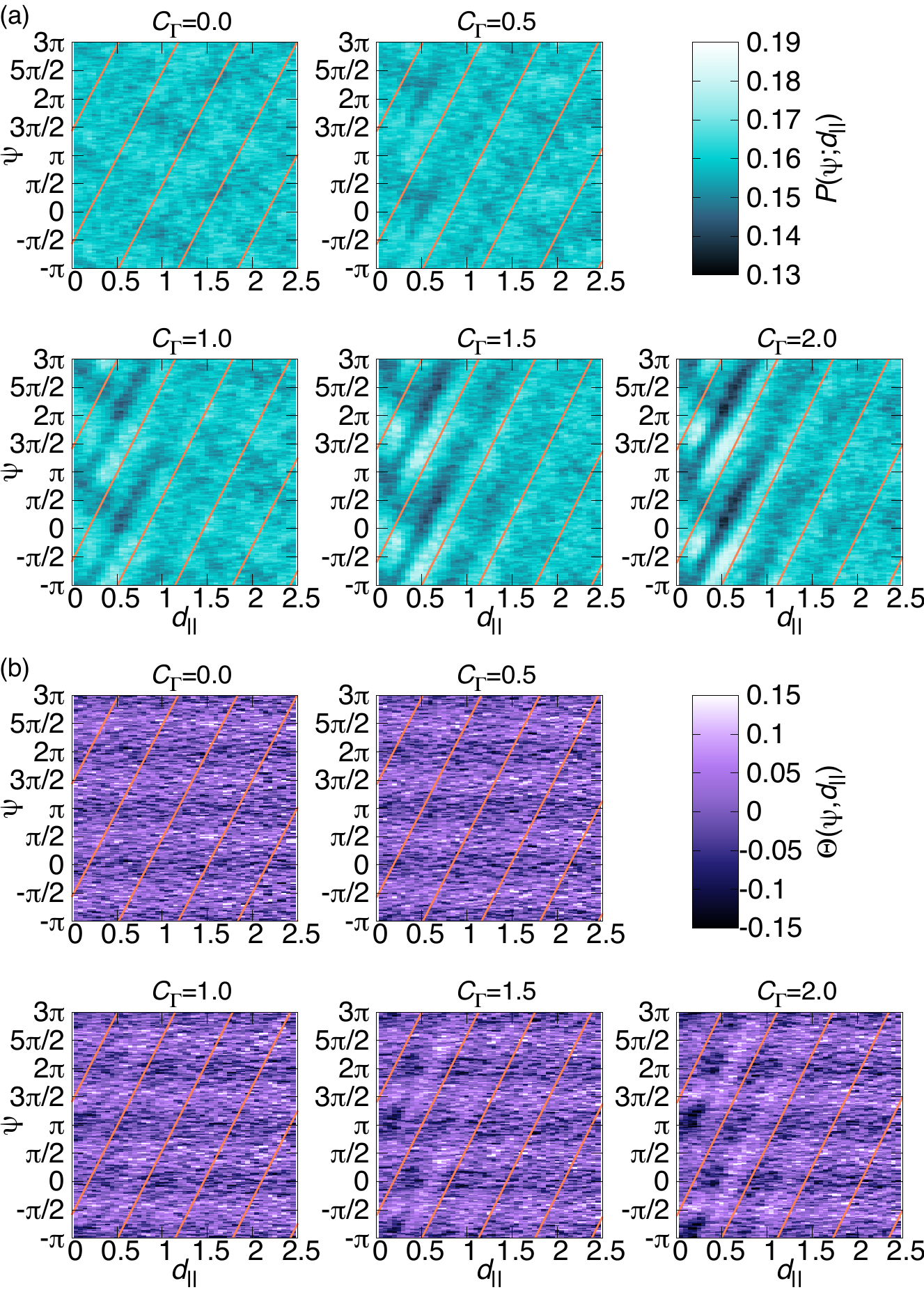}
%bb=0 0 360 252
\caption{
Dependence of (a) the probability distribution $P(\psi;d_\parallel)$ and
(b) the dissipation rate $\Theta(\psi,d_\parallel)$ on $C_\Gamma$ with $d_\perp=0.2$.
Note that plots
in the range $\psi\in[-\pi,\pi]$ is periodically extended to $\psi\in[\pi,3\pi]$ for visibility.
The solid lines represent Eq.~(\ref{THline}) with $\psi_0$ shifted by integer multiples of $2\pi$.
}
\label{PpsiTHTCGam}
\end{figure}

In Fig.~\ref{PpsiTHTCGam}(a), we show $P(\psi;d_\parallel)$ 
for several values of $C_\Gamma$ with $d_\perp = 0.2$ fixed.
For the case without vortices ($C_\Gamma=0$),
the distribution $P(\psi;d_\parallel)$ is almost uniform 
and close to the average value $1/(2\pi) \simeq 0.159$.
Correlation between $\psi$ and $d_\parallel$ 
emerges with increasing $C_\Gamma$,
and a periodic pattern is clearly observed for $C_\Gamma= 2.0$. 
The correlation is strong for the distance $d_\parallel\lesssim1$,
and is detectable up to $d_\parallel \sim2$.
This result is qualitatively the same as the experiment on goldfish~\cite{Li2020} 
(see Fig. 18 of Supplementary Information of the reference).

Theoretically, 
synchronization of the tailbeat is achieved at the
phase difference 
\begin{equation}
\label{THline}
\psi=\frac{2\pi f_a}{U}d_\parallel+\psi_0,
\quad
\psi_0=-\frac{\pi f_a\chi_c}{U}.
\end{equation}
Here, $d_\parallel/U$ gives 
the time for the vortex emitted by the leader to 
travel the distance $d_\parallel$,
and $2\pi f_ad_\parallel/U$ 
is the increment of phase of the leader
in this time interval~\cite{Li2020}.
We formulate the phase shift $\psi_0$ using
the fact that
the vortex emitted from the tip of the leader's plate 
affects the follower most strongly 
at the mid-point of the follower's plate
(see the definitions of the drag force $\bm{F}_d$ and the lift force $\bm{F}_l$ 
in Eqs.~(\ref{Fd}) and~(\ref{Fl})).
This means that the interaction between the two swimmers is strongest 
and their motion is synchronized 
at the distance $d_\parallel=\chi_c/2$,
which gives  $\psi_0=-(2\pi f_a/U)\times(\chi_c/2)$.
As shown in Fig.~\ref{PpsiTHTCGam}(a),
the formula (\ref{THline}) reproduces 
the peak lines of $P(\psi;d_\parallel)$ fairly well.
The phase shift $\psi_0$ is varied by the background flow speed 
$U$ (=$\abs{\delta V_0}$), which is a function of $C_\Gamma$,
and the range is $\psi_0=-\pi\times0.57$-0.61.
The dependence of $P(\psi;d_\parallel)$ on $d_\perp$ is shown in Appendix~\ref{Appenddissipation}.

%//////////////////////////////////////////////////////////////////////////////////////////////////////////////////////////%
\subsection{Spontaneous reduction of energy consumption}

Finally, we study the energy dissipation rate 
%(Eq.~(\ref{Thetadef}))
as a function of the phase difference $\psi$ and the front-back distance $d_\parallel$.
%Here, we trace the swimmer 1 and measure the dissipation rate $\Theta_1(\psi,d_\parallel)$.
%This method of measurement is virtually equal to
%measuring the average of the dissipation rate between the leader and  the follower swimmer.
%Note that we confirmed that the below results are the same as the case
%in which we choose the swimmer 2 for the measurement,
%and do not depend on whether the other swimmer is on the left or right (data not shown).
We traced the swimmer 1 for each run and
take the ensemble average over $10^4$ runs
to define the energy dissipation rate $\Theta(\psi,d_\parallel)$.
Note that this is equivalent to taking the average of the leader and follower
in a single run over a very long time.
%$\Theta_1(\psi,d_\parallel)$
%of $10^4$ simulations in each bin of $\psi$ and $d_\parallel$.

The result is shown in Fig.~\ref{PpsiTHTCGam}(b).
With increasing $C_\Gamma$,
$\Theta(\psi,d_\parallel)$ develops an oblique stripe 
%periodic correlation pattern
that is  parallel to the theoretical line (Eq.~(\ref{THline}))
in the range $d_\parallel\lesssim1$.
This result is consistent with the experimental results on robotic fish~\cite{Li2020}.
%{ここで斜め縞の解釈を述べる。→実験結果（ロボット）と一致。
%位相のずれは難しい→Discussion}
Also, the energy dissipation rate shows 
a periodic dependence on $\psi$ with the period $\pi$,
which is independent of $d_\parallel$
and even without the vortex-mediated interaction  ($C_\Gamma=0$). 
This dependence is explained by
the periodicity of the energy dissipation rate $\Theta(\phi)$ for a solo swimmer
shown in Fig.~\ref{Pphitail}(b);
see Appendix~\ref{Appenddissipation} for a detailed discussion.

\begin{figure}[!t]
\centering
\includegraphics[width=\linewidth]{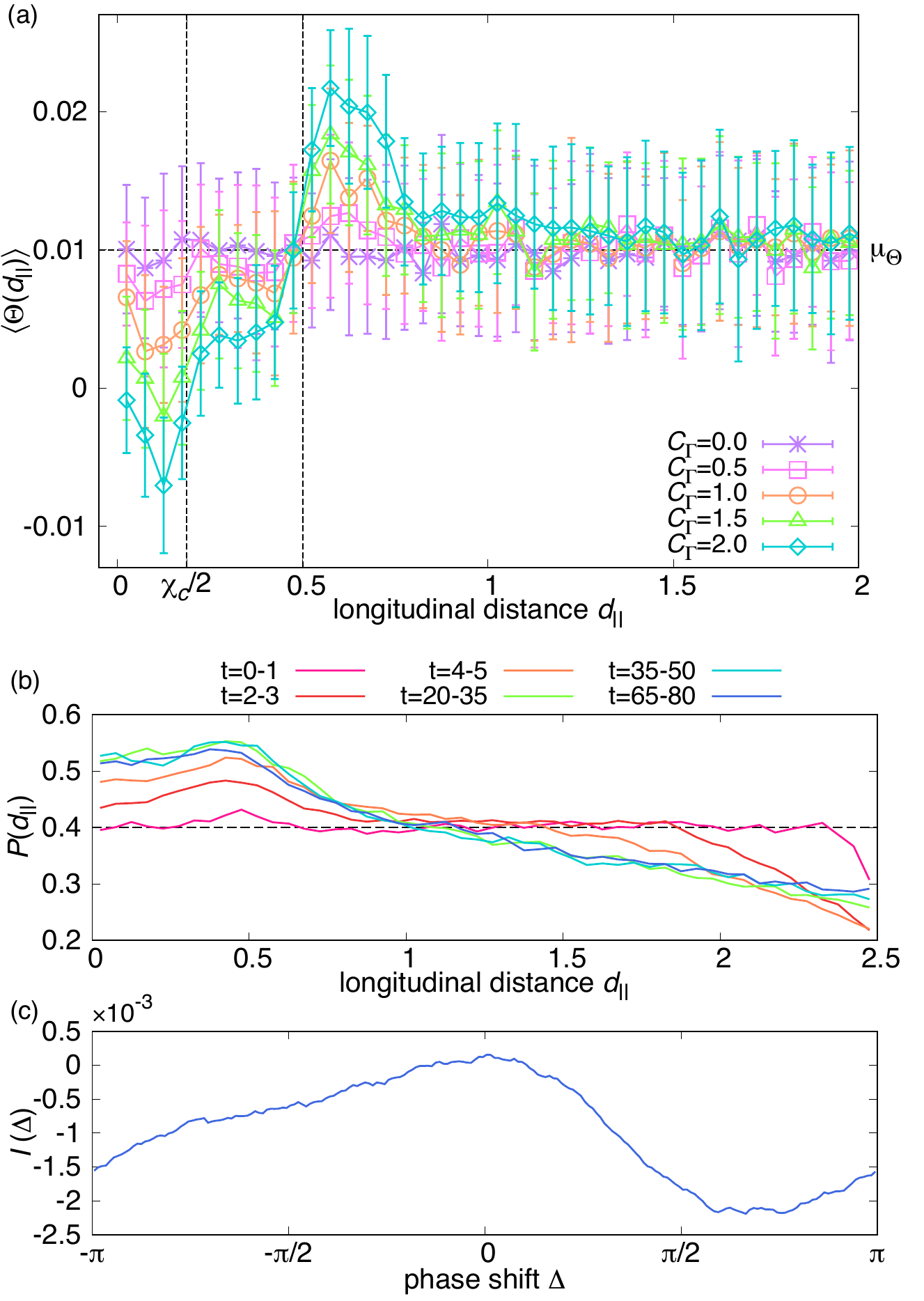}%fig8_d_2_THTave.pdf}%eps}
%bb=0 0 360 252
\caption{
(a) The expected value of the dissipation rate $\langle\Theta\rangle$
as a function of the longitudinal distance $d_\parallel$.
The vortex strength $C_\Gamma$ is varied in $[0,2.0]$.
The error bar is the standard deviation of 50 sets of simulations.
The horizontal dashed line represents the expected value of the dissipation rate $\mu_\Theta$
for solitary swimming.
The vertical dashed lines correspond to a half of the plate length ($\chi_c/2$) and
a half of the body length.
(b) Probability distribution of the longitudinal distance
in different time intervals, normalized in the range $0<d_\parallel < 2.5$.
It develops deviations from the initial uniform 
distribution (horizontal dashed line)  
and reaches a steady distribution with a peak in the range $d_\parallel < 0.5$.
(c) The overlap integral between the probability distribution $P(\psi; d_\parallel)$ 
and the energy dissipation rate $\Theta(\psi, d_\parallel)$ 
as a function of the phase shift $\Delta$
(see text for the definition). 
%, which indicates 
%the degree of deviations from the energetically optimal configuration.
%P(\psi;d_\parallel)\Theta(\psi-\Delta,d_\parallel)
The transverse distance is fixed to $d_\perp = 0.2$ in all the plots 
and $C_\Gamma=2.0$ in (b) and (c).
}
\label{THTaveCGam}
\end{figure}

The expected value of the energy dissipation rate is defined by
\begin{equation}
\langle\Theta(d_\parallel)\rangle=\int_{-\pi}^\pi\dd\psi P(\psi;d_\parallel)\Theta(\psi,d_\parallel).
\end{equation}
We obtain $\langle\Theta(d_\parallel)\rangle$ using 50 sets of $10^4$ simulations
to calculate the average and standard deviation.
%{50万回の平均？エラーバーを出すため}
Fig.~\ref{THTaveCGam}(a) 
shows the dependence of
the averaged $\langle\Theta(d_\parallel)\rangle$ on $C_\Gamma$.
For $C_\Gamma=0$, $\langle\Theta(d_\parallel)\rangle$ 
is independent of $d_\parallel$
and reproduces the value $\mu_\Theta\simeq 0.01$ for solitary swimming, 
as it should
(see also Appendix~\ref{Appenddissipation}).
On the other hand, for $C_\Gamma>0$,
$\Theta$ shows a complex distance dependence.
It makes a minimum in the interval $0 < d_\parallel<\chi_c/2$, 
and even turns negative for $C_\Gamma=1.5$ and $2.0$.
The negative dissipation means that the swimmer gains energy from the flow
in one cycle of tailbeat.
In $\chi_c/2 < d_\parallel < 0.5$,
$\langle\Theta(d_\parallel)\rangle$ shows a gradual increase
and crosses $\mu_\Theta$. 
It makes a sharp peak at  $d_\parallel\sim0.6$,
and converges to $\mu_\Theta$ in $d_\parallel \gtrsim 1$ 
where
the distribution $P(\psi;d_\parallel)$ becomes uniform.
We show the dependence of the energy dissipation rate on $d_\perp$
in Appendix~\ref{Appenddissipation}.

In Fig.~\ref{THTaveCGam}(b), we plot the distribution of 
the longitudinal distance, which is computed in different time intervals
and is normalized in the range $0 < d_\parallel < 2.5$.
It rapidly develops deviations from the initial uniform distribution 
and reaches a steady distribution by $t=20$. 
The steady distribution has a broad peak in the region $d_\parallel < 0.5$,
which corresponds to the region where 
the energy dissipation rate is reduced (see Fig.~\ref{THTaveCGam}(a)).
The distribution decreases with the distance for $d_\parallel > 0.5$.

While we showed that 
the energy dissipation rate is reduced by vortex phase matching,
its minimum is shifted from the most probable phase difference given by Eq.(\ref{THline}),
as shown in Fig.~\ref{PpsiTHTCGam}(b).
In order to quantify the shift,
we define the overlap between the probability distribution 
$P(\psi;d_\parallel)$ and the energy dissipation rate $\Theta(\psi,d_\parallel)$
by
\begin{equation} 
I(\Delta)=\int_{-\pi}^\pi \dd\psi P(\psi;d_\parallel)\Theta(\psi-\Delta,d_\parallel).
\end{equation}
Here, we introduced the phase shift $\Delta$ to find 
the difference between the 
the energetically optimal value of $\psi$ and the most probable value of $\psi$.
If   $I(\Delta)$ is minimized at $\Delta = \Delta_0$,
the former is given by
\begin{equation}
\psi_{E} = \frac{2\pi f_a}{U}d_\parallel + \psi_0 - \Delta_0,
\end{equation}
instead of Eq.(\ref{THline}) for the latter.
In Fig.~\ref{THTaveCGam}(c), we plot the overlap integral 
averaged over the distace $0<d_\parallel<0.5$.
Note that $I(0)$ gives the energy dissipation rate 
in Fig.~\ref{THTaveCGam}(c) averaged over the same distance range.
We find that $I(\Delta)$ is maximal around $\Delta = 0$ and minimal
in the range $\Delta(=\Delta_0)\simeq\pi\times0.6$-$0.7$.
It indicates that the actual distribution of the phase difference 
does {\it not} minimize the energy dissipation rate.

%%%%%%%%%%%%%%%%%%%%%%%%%%%%%%%%%%%%%%%%%%%%%%%%%%%%%%%%%%%%
\section{Conclusions}

We constructed a new self-propelled model 
that reproduces many experimental features of 
carangiform and subcarangiform swimming.
For the hydrodynamic part, 
we adopted the quasi-steady approximation~\cite{Nagai1996,Hirayama2000}
and introduced the Rankine vortex street 
which was not considered in the previous 
self-propelled models~\cite{Tchieu2012,Gazzola2016,
Filella2018,Deng2021,Gazzola2015}.
By incorporating the physiological noises, 
we modeled time evolution of the phase of tailbeat and
the distance of swimmers,
which allowed the model fish to
spontaneously select the swimming pattern.
This is regarded as a significant advance from 
the previous models that fix the phase of tailbeat (and the relative distance)~\cite{Hemelrijk2015,Daghooghi2015,Maertens2017,
Li2019,Zhu2014,Park2018,Peng2018,
Dewey2014,Boschitsch2014,Becker2015,Newbolt2019,Oza2019}.
For body kinematics, our model with a single flapping plate
is simpler than the two plates model~\cite{Nagai1996, Hirayama2000}
and the elastic plate models~\cite{Taylor1952,Gazzola2015},
and has the merit to reduce computational cost.
Now let us discuss the type of fish and swimming mode 
for which our model can be applied.
As we take the body length as the unit of length, 
our model can be applied to a (sub)carangiform swimmer with any body length, 
30 to 50 percents of which exhibits undulating motion.
We choose the hydrodynamic parameters so that $\HyRe\sim\mathcal{O}(10^5)$,
which corresponds to the steady swimming of a typical (sub)carangiform swimmer
(see also Appendix~\ref{AppendExperi} for the choice of parameters).
In the steady swimming,
the swimming speed 1-5 BL/s is maintained for more than 1 hour.
On the other hand,  our model cannot be applied to 
a swimming speed $\gtrsim5$ BL/s, which corresponds to the fast-start
with the duration $\sim0.1$ sec and $\HyRe\gtrsim\mathcal{O}(10^6)$~\cite{Domenici1997}.
In addition, we do not consider burst-and-coast swimming 
which has a cycle with active phase and gliding phase
instead of the continuous fin motion~\cite{GLi2021}.
The burst-and-coast swimming is typically adopted by small fish 
and gives $\HyRe\lesssim\mathcal{O}(10^4)$. 

For solo swimming,
%we confirmed that our model nicely reproduces 
%the experimental results First, 
the model reproduces 
the linear relation between the frequency of tailbeat and the thrust speed 
established for many species of fish~\cite{Tanaka1996,Nagai1979,Bainbridge1958,Akanyeti2017} 
(Fig.~\ref{Vffa}).
%of the   (sub)carangiform
Self-induced vortices have only a minor effect on the thrust speed,
as found with the elastic plate model~\cite{Gazzola2015}.
The elastic plate model also reproduced the frequency-speed relation,
but by assuming the amplitude of tailbeat an order of magnitude smaller 
than the typical experimental value $A_0=0.1 L_b$. 
It implies that the elastic plate swimmer 
has much higher swimming efficiency than real fish.
We used the standard amplitude $A_0/L_b=0.1$ 
and the plate length $\chi_c=0.375$ to reproduce the relation.
In addition, our model reproduces the typical 
Strouhal number $\HySt\approx0.3$~\cite{Gazzola2014,Triantafyllou1993,Taylor2003}
(see Fig.~\ref{StVorb}(a)-(c)),
and the trajectory of tailbeat agrees with 
that of dace in steady swimming~\cite{Bainbridge1963} (see Fig.~\ref{StVorb}(d)).

The two noise amplitudes ($D_a, D_\varphi$) 
enabled us to fine-tune the probability distributions of the thrust speed 
and the amplitude and frequency of the tailbeat.
We chose $D_a=0.7$ and $D_\varphi=0.25$
by matching the distribution functions 
with those of goldfish~\cite{Li2020}
(see Fig.~\ref{tevdis}(b)-(d) and Appendix~\ref{Appenddistri}).
Note that the physiological noises affect the motion of the caudal muscles,
and therefore enter the equations of motion only via the torque.
We neglected the effect of hydrodynamic turbulence,
which could be introduced in the model as noises in both the thrust force and torque.
The physiological and hydrodynamic noises may have different roles 
in modifying the swimming pattern, 
but they are beyond the scope of the present study. 

For a pair of swimmers, we showed statistically
that they adjust the phase difference 
depending on the front-back distance.
The probability distribution of the phase difference 
$\psi$ shows a strong anisotropy
in the short distance $d_\parallel\lesssim1$ (see Fig.~\ref{PpsiTHTCGam}(a))
which is qualitatively consistent with the result on goldfish~\cite{Li2020}.
The correlation between $\psi$ and $d_\parallel$
is described by the theoretical relation (\ref{THline}),
where we explicitly derived the formula for the phase shift $\psi_0$.
The formula is generalized as $\psi_0 = -2\pi f_a d_0/U$, 
where $d_0$ is the distance 
between the plate tip and 
the point on the plate P 
where the flow has the strongest effect on the flapping motion.
We numerically obtained 
$\psi_0\approx-0.6\pi$,
while the experiment on goldfish shows $\psi_0\approx-0.2\pi$.
The difference between the two results could be explained as follows.
In our model, P is the mid-point of the plate ($d_0 = l_c/2$).
On the other hand, the experimental value of $\psi_0$
implies that the point P is closer to the tip of the fin than in our model.
This seems to be consistent with fact that 
the stiffness of the caudal part decreases
as we go from the peduncle to the tip of the fin~\cite{McHenry1995},
and thus the fin is easier to be deformed than the peduncle.

The energy dissipation rate $\Theta(\psi,d_\parallel)$ also 
has a pattern described by 
the linear relation between $\psi$ and $d_\parallel$
in $d_\parallel\lesssim1$ (see Fig.~\ref{PpsiTHTCGam}(b)).
However, the phase difference $\psi_E$ 
that minimizes $\Theta(\psi,d_\parallel)$ 
is shifted from the most probable phase difference
by $-\Delta_0$.  
The value of $\psi_E$ at $d_\parallel =0$ is given by
$\psi_{E0} = \psi_0 - \Delta_0$,
and the numerically obtained values 
$\psi_0\approx -0.6\pi$ and $\Delta_0\approx \pi\times 0.6-0.7$
give $\psi_{E0}\approx \pi\times 0.7-0.8$ (mod $2\pi$).
%The physical origin of the shift is unclear, 
%but a similar tendency 
A shift between 
the energetically optimal and the most probable phase differences
is also found in the experiment~\cite{Li2020}.
There, the power efficiency of a robotic fish 
is fitted by the same formula with $\psi_{E0}\approx0.3\pi$~\cite{Li2020},
and is larger than $\psi_0$ for the distribution of goldfish 
by $0.5\pi$. 
The origin of the difference between 
our and experimental values of $\psi_{E0}$ is unclear, 
but we should note that the probability distribution for 
robotic fish may differ from that of goldfish,
and it could explain the difference.
More importantly, the nonzero value of $\Delta_0$
means that the realized phase difference is not the one
that minimizes the energy dissipation rate.
This is in line with the observation 
%in other systems
that hydrodynamic synchronization
does not always minimize energy dissipation rate~\cite{Elfring2009,Liao2021},
and supports the importance to model the phase dynamics.

The experiment~\cite{Li2020} also showed that
the change in the power efficiency is small
in the range of the transverse distance of 0.27-0.33 body length,
which is also reproduced by our model
(see $d_\perp=0.25$-0.3 in Fig.~\ref{PpsiTHTdperp}(b)).
On the other hand, the horizontal stripe pattern in 
Fig.~\ref{PpsiTHTCGam}(b) is explained analytically
by the distributions for a solo swimmer;
see Appendix~\ref{Appenddissipation}.
This pattern is not 
found in the experimental result on robotic fish, 
possibly due to the short measurement range of 
the longitudinal distance %(0-1 body length) 
and large fluctuations~\cite{Li2020}.

The expected energy dissipation rate $\langle\Theta(d_\parallel)\rangle$
shows that 
the spontaneous reduction of energy consumption 
for the distance $d_\parallel\lesssim0.5$,
which roughly corresponds to the region of
the strongest hydrodynamic interactions for a pair of goldfish~\cite{Li2020}. 
At a larger distance, the energy dissipation rate
becomes larger than that of solo swimming.
These results are qualitatively consistent with 
those for robotic fish~\cite{Li2020}:
In Fig. 10 of Supplementary Information of the paper, 
it is shown that the efficiency of the electric power increases
when the longitudinal distance is less than 0.7 body lengths,
and can adopt negative value at the distance $\sim0.8$ body lengths.
Furthermore, we found that the distribution of the longitudinal distance
develops a peak in the region $d_\parallel < 0.5$ in the course of time.
It means that the energetically favorable distance is dynamically selected by 
the swimmers.
%With these results, we conclude that our model  
%reproduced
%the experimentally observed vortex phase matching and 
%spontaneous reduction of energy consumption for the first time.

Finally, there are some aspects to be addressed in the future.
%We can apply the model to any fish with subcarangiform and carangiform
%by tuning the ratio $l_c/L_b$.
%%% carangiform は寸胴型でマグロ型に比べるとしっぽが長方形に近い
Our model can be applied to carangiform and subcarangiform swimmers, 
but not to the other types 
of fish~\cite{Sfakiotakis1999,Lauder2005}.
To reproduce the anguilliform, which is an undulating motion of the large part of the body,
we need to introduce more hinges such as in the elastic plate model~\cite{Gazzola2015}.
The shape of the caudal fin is also an important factor to determine
the swimming characteristics~\cite{Hirayama2000}.
The rectangular plate in our model is suited to 
many of carangiform and subcarangiform swimmers~\cite{Bainbridge1963}, 
while tuna, representing the thunniform,
%that uses only 20-30 \% of the body length for locomotion,
has a thin crescent-shaped caudal fin.
The thrust speed we obtained for $\chi_c=0.3$ is smaller than 
the measurement for
yellowfin tuna (\textit{Thunnus albacares})~\cite{Shadwick2008} 
(see also Fig.~\ref{StVorb}(c)),
which is possibly due to the difference in the fin shapes.

The Rankine vortex street gives a fair representation of the vortex flow 
at low computational cost,  but the flow field around real fish 
is more complex.
For example, we may incorporate the dipolar flow field~\cite{Filella2018}
and collision process between the vortex and the swimmers~\cite{Hemelrijk2015}.
Turbulent flow at a small scale might be treated as noises 
on the thrust force and torque, although their mode structures are highly nontrivial.
Our single-hinge model is a minimal model of locomotion gait,
and does not capture the motion of the anterior part of the body.
%We assume that the anterior part of body is rigid.
%Actually, 
Although the head-beat amplitude is typically an order of magnitude 
%0.1 times as small as 
smaller than
the tailbeat amplitude~\cite{Santo2021},
%the movement of the anterior part
%however, 
it can generate a sizable thrust force %on near the head 
in the anterior part~\cite{Wen2013,Lucas2020}.
%by the negative pressure
This thrust force affects
%which related with the 
hydrodynamic interaction in in-line configurations~\cite{Thandiackal2023}.
The present work focuses on vortex phase matching 
at transversal distance $d_\perp \gtrsim 0.2$ and $d_\parallel\lesssim 1$,
and the analysis of vortex-mediated interaction
in in-line configurations with $d_\perp \sim 0$ and $d_\parallel\gtrsim 1$ 
is left for future work.
%($d_\perp \gtrsim 0.2$ and $d_\parallel\gtrsim1$.
%Note that, although introducing the movement and the thrust force of the anterior part
%changes the our results quantitatively
%(e.g. the energy consumption at the $d_\parallel\gtrsim1$),
%the periodic pattern as shown in Fig.~\ref{PpsiTHTCGam} does not change qualitatively
%because the interaction related with the thrust force of the anterior part is
%uncorrelated with phase difference $\psi$.
Our model is also limited to one-dimensional motion
mimicking the experiment with background flow
(for example, in the experiment with two gold fish~\cite{Li2020},
they move in the lateral direction only in the range 0.27-0.33 BL).
In order to extend it to two- or three-dimensional motion,
we would need to integrate it with phenomenological self-propelled particle models 
with repulsion, attraction, and alignment interactions,
which are presumably topological~\cite{
Gautrais2012, Calovi2014, Filella2018, Ito2022a, Ito2022b},
and also the three-dimensional flow field of a vortex ring~\cite{Nauen2002}.
Inclusion of these aspects will be an interesting issue for the future.

%%%%%%%%%%%%%%%%%%%%%%%%%%%%%%%%%%%%%%%%%%%%%%%%%%%%%%%%%%%%
\acknowledgments
This work was supported by a research environment of Tohoku University,
Division for Interdisciplinary Advanced Research and Education.
We acknowledge financial support by JSPS KAKENHI Grant Number 23KJ0171 to Susumu Ito.

%%%%%%%%%%%%%%%%%%%%%%%%%%%%%%%%%%%%%%%%%%%%%%%%%%%%%%%%%%%%
\section*{Author Contributions}
Susumu Ito: Conceptualization (lead); Investigation (lead); Validation (lead); Visualization (lead); Writing - original draft (lead); Writing - review \& editing (equal). 
Nariya Uchida: Conceptualization (supporting); Supervision (lead); Writing - review \& editing (equal).

%%%%%%%%%%%%%%%%%%%%%%%%%%%%%%%%%%%%%%%%%%%%%%%%%%%%%%%%%%%%
\appendix
\section{Hilbert transformation}
\label{AppendHilbert}

\begin{figure}[!b]
\centering
\includegraphics[width=\linewidth]{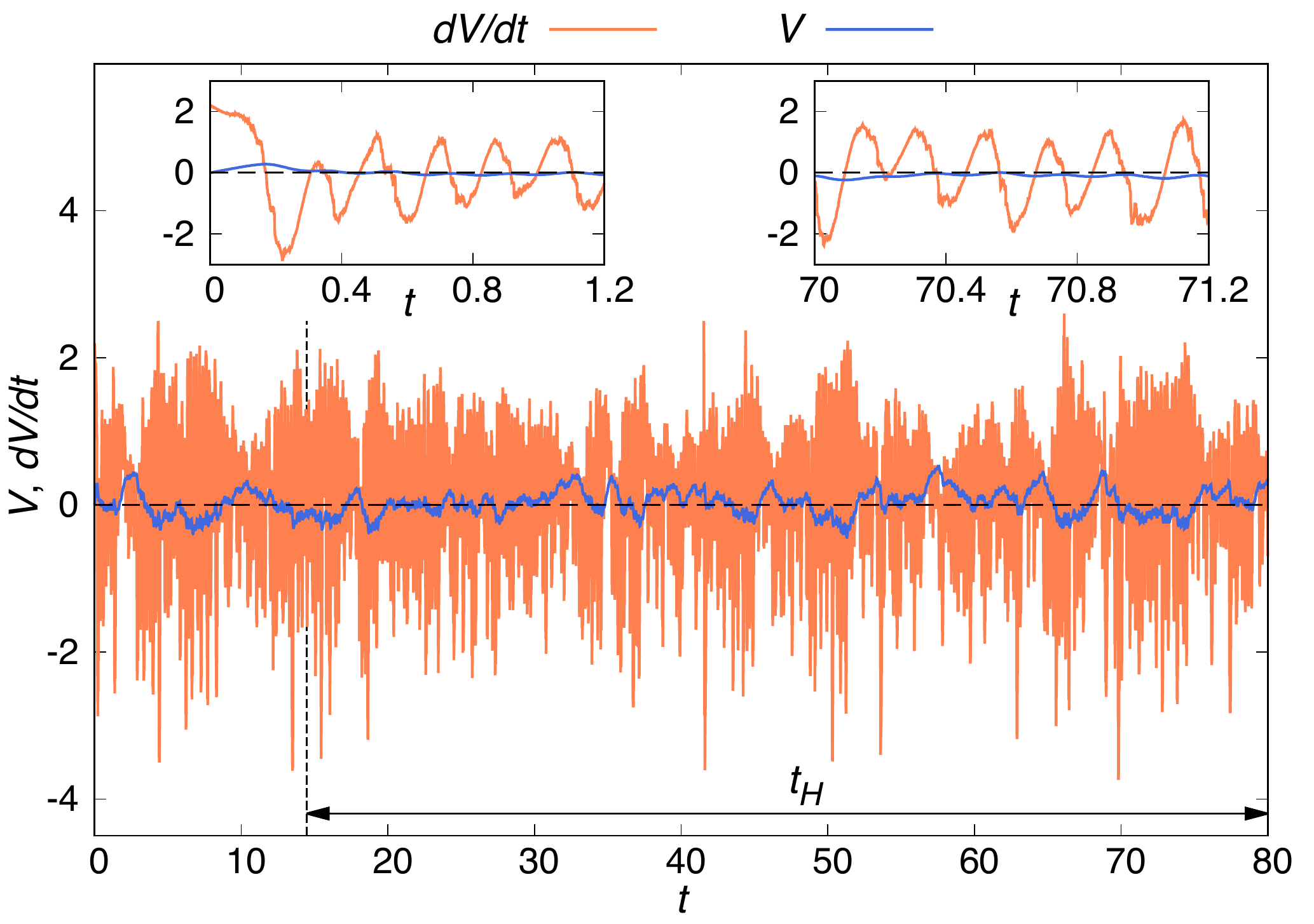}
%bb=0 0 360 252
\caption{
Time evolution of the acceleration $\dv*{V}{t}$ and the velocity $V$ of a swimmer
from the initial state to the end of a simulation ($t=80$)
with $\chi_c=0.375,~f_a=2.5,~C_\Gamma=2.0,~D_a=0.7$, and $D_\varphi=0.25$.
The Hilbert transform is conducted in the time window with the duration $t_H$.
The left top inset represents the time evolution for $t\in[0,1.2]$,
and the right top inset represents the time evolution for $t\in[70,71.2]$.
}
\label{tevHilexam}
\end{figure}

The Hilbert transformation is a mathematical method 
to extract the amplitude and the phase from
oscillating time series data~\cite{Huang2014}.
We define the Hilbert transform $\hat{y}_c(t)$ of the plate tip motion $y_c(t)$
by the principal value integral
\begin{equation}
\label{Pint}
\hat{y}_c(t)=\frac{1}{\pi}\mathcal{P}\int_{-\infty}^\infty\dd t'\frac{y_c(t')}{t-t'}.
\end{equation}
Using the Fourier representation
\begin{equation}
y_c(t)=\frac{1}{2\pi}\int_{-\infty}^\infty\dd\omega \, y_{c,\omega}e^{i\omega t}
\end{equation}
and residue theorem of complex integtation,
the integral~(\ref{Pint}) becomes
\begin{equation}
\label{Hilbertdef}
\hat{y}_c(t)=\frac{1}{2\pi}\int_{-\infty}^\infty\dd\omega\,
y_{c,\omega}\sgn{\omega}e^{i\leri{\omega t-\frac{\pi}{2}}},
\end{equation}
where $\sgn{\circ}$ is the sign function (see Eq.~(\ref{sgnfunc})).
For example, for constants $a$, $\omega$, and $\phi_0$,
$y_c(t)=a\cos(\omega t+\phi_0)$ and $y_c(t)=a\sin(\abs{\omega}t+\phi_0)$ is transformed to
$\hat{y}_c(t)=a\sin(\abs{\omega}t+\phi_0)$ and $\hat{y}_c(t)=a\cos(\omega t+\phi_0)$, respectively.
Based on the formula~(\ref{Hilbertdef}), we can define the amplitude
\begin{equation}
A(t)=\sqrt{y_c^2(t)+\hat{y}_c^2(t)}
\end{equation}
and the phase
\begin{equation}
\phi(t)=\tan[-1](\frac{\hat{y}_c(t)}{y_c(t)}).
\end{equation}
This definition indicates $y_c(t)=A(t)\cos\phi(t)$.

Fig.~\ref{tevHilexam} shows time evolution of the acceleration $\dv*{V}{t}$ and the velocity $V$ of a swimmer.
As indicated in the left top inset, the acceleration establishes a steady oscillatory pattern
in a short time $t\sim {\cal O}(1)$,
and we conducted the Hilbert transformation for the steady swimming.

%%%%%%%%%%%%%%%%%%%%%%%%%%%%%%%%%%%%%%%%%%%%%%%%%%%%%%%%%%%%
\section{Drag and lift coefficient $C_d,~C_l$ and coefficient $K$ of added mass}
\label{AppendCdClK}

\begin{figure}[!b]
\centering
\includegraphics[width=\linewidth]{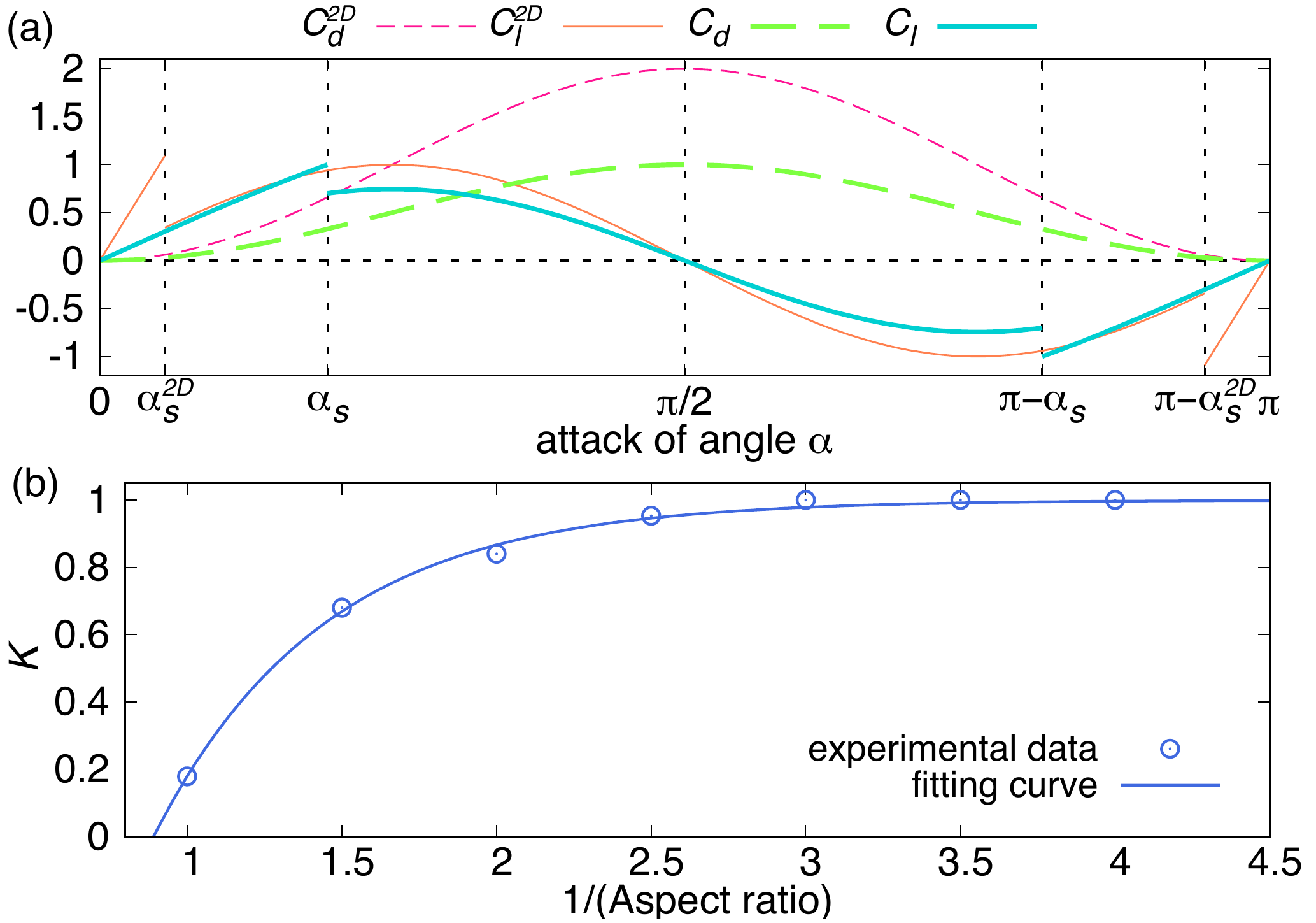}%eps}
%bb=0 0 360 252
\caption{
Plots of the hydrodynamic coefficients.
(a) Dependences of $C_d$ (thick dashed line) and $C_l$ (thick solid line)
on the angle of attack $\alpha$.
The parameters are $c_d=1.0,~c_{l0}=1.0,~c_l=0.7,~\alpha_s=35^\circ$.
$C_d^{2D}$ (thin dashed line) and $C_l^{2D}$ (thin solid line) are coefficients
for a two dimensional plate aerofoil ($\alpha_s^{2D}=10^\circ$).
(b) The added mass prefactor $K$ as a function of the inverse aspect ratio.
The points show the experimental data~\cite{Brennen1982}.
The solid line fits the data by Eq.~(\ref{Kfit}) with
$a_K=1.82,~b_K=0.89$.
}
\label{CdClKfig}
\end{figure}

The drag coefficient $C_d(\alpha)$ and the lift coefficient $C_l(\alpha)$ of
a plate with a finite aspect ratio in Eqs.~(\ref{Fd}) and (\ref{Fl})
are derived by modifying the formula
for a two-dimensional plate~\cite{Ortiz2015}.
The two-dimensional plate has the drag coefficient
\begin{equation}
C_d^{2D}(\alpha)=2\sin^2\alpha,
\end{equation}
and the lift coefficient
\begin{align}
C_l^{2D}(\alpha)=\left\{ \begin{array}{ll}
2\pi\sin\alpha&[\alpha<\alpha_s^{2D}],\\
\sin2\alpha&[\alpha_s^{2D}<\alpha<\pi-\alpha_s^{2D}],\\
-2\pi\sin\alpha&[\alpha>\pi-\alpha_s^{2D}],
\end{array} \right.
\end{align}
where $\alpha_s^{2D}\sim10^\circ$ is a stall angle
due to separation of flow at the rear of the airfoil
(see Fig.~\ref{CdClKfig}(a))~\cite{Jiang2014}.
These coefficients are symmetric about $\alpha=\pi/2$.
For a plate of finite aspect ratio,
$C_d(\alpha)$ and $C_l(\alpha)$ become smaller than
$C_d^{2D}(\alpha)$ and $C_l^{2D}(\alpha)$, respectively,
and a stall angle $\alpha_s$ is larger than $\alpha_s^{2D}$~\cite{Ortiz2015},
because the pressure is dispersed 
in the transverse direction and prevents flow separation.
Referring to the data in Ref.~\cite{Ortiz2015},
we formulate the drag coefficient as
\begin{equation}
\label{Cd}
C_d(\alpha)=c_d\sin^2\alpha,
\end{equation}
and the lift coefficient as
\begin{align}
\label{Cl}
C_l(\alpha)=\left\{ \begin{array}{ll}
c_{l0}\frac{\sin\alpha}{\sin\alpha_s}&[\alpha<\alpha_s],\\
c_l\frac{\sin2\alpha}{\sin2\alpha_s}&[\alpha_s<\alpha<\pi-\alpha_s],\\
-c_{l0}\frac{\sin\alpha}{\sin\alpha_s}&[\alpha>\pi-\alpha_s].
\end{array} \right.
\end{align}
(see Table~\ref{experimentalvalue} for the values of $c_d,~c_{l0},~c_l,~\alpha_s$).
They are plotted and compared with the two-dimensional case
in Fig.~\ref{CdClKfig}(a).

Next, we consider the added mass of a plate in Eq.~(\ref{addedmass}).
An added mass per unit length of an oscillating two-dimensional plate
with height $H_b$
in a potential flow is theoretically given by~\cite{Wendel1956}
\begin{equation}
m_p^{2D}=\frac{\pi}{4}\rho_wH_b^2.
\end{equation}
For a plate with a finite aspect ratio ($H_b/l_c\lesssim1$) 
in Eq.~(\ref{addedmass}),
we have the additional factor $K (\le 1)$
that depends on the inverse aspect ratio $x = l_c/H_b$.
The behavior of $K(x)$ is determined using the experimental results~\cite{Brennen1982},
and $K\rightarrow1$ in the low aspect ratio limit;
see Fig.~\ref{CdClKfig}(b).
Because $K$ rapidly approaches 1 as $x \to \infty$, 
we 
assumed the exponential function
\begin{equation}
\label{Kfit}
K(x)=1-e^{-a_K(x -b_K)}
\end{equation}
and determined the constants $a_K$ and $b_K$ by 
fitting to the experimental data. 

%\begin{table}[!t]
%\begin{center}
%\caption{Dependences of $K$ on $\chi_c$.}
%\begin{tabular}{c||c|c|c|c|c|c|c}
%$\chi_c$&0.300&0.325&0.350&0.375&0.400&0.425&0.450\\\hline
%$K$&0.18&0.30&0.40&0.48&0.55&0.62&0.67\\
%\end{tabular}
%\label{Ktable}
%\end{center}
%\end{table}

%//////////////////////////////////////////////////////////////////////////////////////////////////////////////////////////%
\section{Parameters}
\label{AppendExperi}

\begin{table*}[!t]
\begin{center}
\caption{
List of the parameters.
We rescaled the parameter values by
the body length $L_b $, timescale $\tau_0=1$sec, and body mass $M$.
(BL means the body length.)
}
\begin{tabular}{cccc}
symbol& meaning & experiment & simulation (rescaled)\\\hline
$H_b$ & body height & $\sim0.3$ BL~\cite{Jones1999} & --\\
$l_c$&length of the caudal part&$\sim0.3$-0.5 BL~\cite{Lauder2005,Sfakiotakis1999}&--\\
$\chi_h$&$H_b/L_b$&--&0.3\\
$\chi_c$&$l_c/L_b$&--&0.3-0.45\\
$A_0$&amplitude of tailbeat&$\sim0.1$ BL~\cite{Bainbridge1958,Hunter1971,Webb1984,Akanyeti2017,Li2021}&0.1\\
$\rho$&effective density of the body&$\sim41$ kg m$^{-3}$~\cite{Jones1999}&--\\
$\rho_w$&density of water&$\sim1000$ kg m$^{-3}$&--\\
$\chi_\rho$&$\rho_w/\rho$&--&25\\
$C_D$&drag coefficient of the body&$\sim0.02$-0.07~\cite{Tandler2019}&0.037~\cite{Ilio2018}\\
$c_d$&drag coefficient of the plate&$\sim1.0$~\cite{Ortiz2015}&1.0\\
$c_{l0}$&lift coefficient of the plate (before stall)&$\sim1.0$~\cite{Ortiz2015}&1.0\\
$c_l$&lift coefficient of plate the (after stall)&$\sim0.7$~\cite{Ortiz2015}&0.7\\
$\alpha_s$&stall angle&$\sim35^\circ$~\cite{Ortiz2015}&35$^\circ$\\
$r_R$&radius of a vortex core&$\sim0.04$-0.05 BL~\cite{Nauen2002,Akanyeti2017}&0.04\\
$\Gamma$&circulation of a vortex&$\sim0.06$-0.25 BL$^2$ s$^{-1}$~\cite{Nauen2002,Wise2018}&--\\
$C_\Gamma$&prefactor of the circulation&$\gtrsim1.0$~\cite{Schnipper2009,Agre2016}&0.0-2.0\\
$\tau_\Gamma$&decay time of a vortex&$\sim2.0$ s~\cite{Oza2019}&2.0\\
$B$&bending stiffness&$\sim10^{-4}$ N m$^2$~\cite{McHenry1995}&1.0\\
$f_a$&active frequency&--&1.0-7.5\\
$\tau_a$&damping timescale of $N_a$&--&1.0\\
$D_a$&diffusion coefficient for $N_a$&--&0-1.0\\
$D_\phi$&diffusion coefficient for $\varphi$&--&0-1.0\\
$d_\perp$&transverse distance&0.27-0.33 BL~\cite{Li2020}&0.2-0.4\\\hline
\end{tabular}
\label{experimentalvalue}
\end{center}
\end{table*}

Here, we define the parameters in our model 
by comparison to experimental values. 
See Table~\ref{experimentalvalue} for summary of the parameters.

We use the body height $H_b=0.3$ BL (body length)
which is the averaged value of many species of fish~\cite{Jones1999},
and the length of caudal part $l_c=0.3$-0.45 BL 
for carangiform and subcarangiform~\cite{Lauder2005,Sfakiotakis1999}.
The effective density of body $\rho$ is 41 kg m$^{-3}$
is obtained by averaging over various species of fish~\cite{Jones1999},
and thus we use  $\chi_\rho=\rho_w/\rho=25$.
The amplitude of the tailbeat $A_0=0.1$ BL
is ubiquitous for many species of fish at various swimming speeds~\cite{Bainbridge1958,Hunter1971,Webb1984,Akanyeti2017,Li2021,NoteA0}.

In our system,
the speed of background flow is $U/L_b\sim\mathcal{O}(1)$ s$^{-1}$,
the fish size is $L_b\sim\mathcal{O}(0.1)$-$\mathcal{O}(1)$ m,
and the kinematic viscosity of water is on the order of $\nu\sim\mathcal{O}(10^{-6})$ m$^2$/s.
Therefore, the Reynolds number is $\HyRe=UL_b/\nu\sim\mathcal{O}(10^5)$,
which corresponds to the Reynolds number of many species of fish~\cite{Gazzola2014}.
We adopt the hydrodynamic parameters from the experiments
to reproduce the Reynolds number $\HyRe\sim\mathcal{O}(10^5)$.
The drag coefficient of the body $C_D=0.037$ is 
that of the two-dimensional NACA0012 airfoil
at Reynolds number $\HyRe\sim10^5$~\cite{Ilio2018}.
The NACA0012 airfoil is often used
as a substitute for a fish body~\cite{Maertens2017}.
This value of $C_D$ is also close to that of a dead fish 
at $\HyRe\sim\mathcal{O}(10^4)$-$\mathcal{O}(10^5)$:
bluegill (\textit{Lepomis macrochirus}), rainbow trout (\textit{Oncorhynchus mykiss}),
and zebrafish (\textit{Danio rerio})~\cite{Tandler2019}.
The parameters $c_d$, $c_{l0}$, $c_l$, and $\alpha_s$ in
the drag coefficient of a plate $C_d(\alpha)$ (Eq.~(\ref{Cd}))
and the lift coefficient $C_l(\alpha)$ (Eq.~(\ref{Cl})) are read from 
the data of plates with a low aspect ratio $\lesssim1$ with $\HyRe\sim10^5$
(Fig. 4(a)-(b) in Ref.~\cite{Ortiz2015}).

The radius of the vortex core is set to $r_R=0.04$ BL using 
the data for the (sub)carangiform swimmers
(rainbow trout~\cite{Akanyeti2017}
and  chub mackerel (\textit{Scomber japonicus})~\cite{Nauen2002})
in steady swimming.
We can estimate the circulation of a vortex as
$\Gamma\sim0.06$-0.25 BL$^2$ s$^{-1}$
for (sub)carangiform fish (bluegill~\cite{Wise2018} and chub mackerel~\cite{Nauen2002}).
For the frequency $f_a=2.5$ s$^{-1}$, which the reference value in our simulations,
the observed values of $\Gamma$ correspond to 
$C_\Gamma=0.5$-2.0 in Eq.~(\ref{Gamma}).
We set $\tau_\Gamma=2.0$ s by virtue of Supplemental Material in Ref.~\cite{Oza2019},
where the typical value of $\tau_\Gamma$ is estimated from 
some experimental results on the vortex street.

The bending stiffness $B$ is on the order of $10^{-4}$~N~$\cdot$~m$^2$
around the caudal peduncle of pumpkinseed sunfish (\textit{Lepomis gibbosus})~\cite{McHenry1995}.
%{Fig.2 of the paper. }
Non-dimensionalization with the body length $L_b=0.13$ m of pumpkinseed sunfish~\cite{McHenry1995}
and the body mass formula $M=\rho\chi_hL_b^3$ (Eq.~(\ref{mass})) gives $B\sim1.0$. 
The adopted range of the active frequency $f_a$ 
correspond to that of the steady swimming~\cite{Wu1977}:
when a fish swims with sustained or prolonged activity performance
which are maintained for indefinitely and 1 to 2 hours, respectively,
the swimming speed is 1-5 BL/s which corresponds to $f_a=$1.0-7.5 s$^{-1}$ (see Fig.~\ref{Vffa}).
The timescale $\tau_a=1.0$ is the same as the timescale of velocity change
estimated by steady swimming of fish~\cite{Ito2022a,Ito2022b}:
the estimated timescale of velocity change is the same as $\tau_0$,
and it is reasonable that the timescale of muscle activity
change is almost the same as the estimated timescale of velocity change ($\tau_a\simeq\tau_0=1.0$).
On the other hand, we tuned the noise amplitudes $D_a$ and $D_\varphi$
by fitting the distributions of the thrust velocity $V$, tailbeat amplitude $A$, 
and frequency $f$ with the experimental data~\cite{Li2020}.
In our phenomenological model of CPG,  it is not possible to 
estimate $D_a$ and $D_\varphi$ by comparison 
with the electromyography data.
%is determined qualitatively
%due to the phenomenologically physiological noise.
If we use more detailed neural circuits/CPG models~\cite{Matsuoka2011},
it may be possible to determine the noise parameters quantitatively.
We adjust the transverse distance $d_\perp$ between a pair of swimmers 
to cover the experimental conditions~\cite{Li2020}.

Finally, to clarify the hydrodynamic performance of a swimmer,
we estimate the magnitude of the hydrodynamic forces.
In the following, we use the non-dimensional force
with the unit of force $F_0=ML_b/\tau_0^2$, which
can be evaluated from the body mass $M$ and body length $L_b$ and
$\tau_0 \sim 1$.
%gives a body mass an acceleration of a body length per second per second.
%We can assign $M$ and $L_b$ of 
%can be evaluated for a (sub)carangiform fish 
%with $\HyRe\sim\mathcal{O}(10^5)$
%and obtain $F_0$ for the fish.
% In the following, we use the non-dimensional force $F/F_0$,
% 
% and we can immediately back to the force with dimension by considering the product with $F_0$.
%
%In addition, 
We neglect the vortex flow because the effect of the self-induced vortex on a swimmer
is small as shown in Fig.~\ref{Vffa}.
The drag force on the body is estimated as $F_D\sim\chi_\rho C_D\delta V^2$.
The thrust speed $\delta V$ is about 1.5 with $f_a=2.5$ (see Fig.~\ref{Vffa}),
and then we obtain $F_D\sim2.1$.
The drag force and the lift force on the plate is
$F_d\sim0.5\chi_\rho\chi_cC_d(\alpha)\cos\beta(\bm{v}+\delta\bm{V})^2$
and $F_l\sim0.5\chi_\rho\chi_cC_l(\alpha)\sin\beta(\bm{v}+\delta\bm{V})^2$, respectively.
$\alpha$ and $\beta$  depends on time and the angle of the plate $\theta$.
We estimate $F_d$ and $F_l$ only at $\theta=0$ where
$\abs{\bm{v}}$ has the maximum value in the periodic fin motion (see Fig.~\ref{StVorb}):
$\abs{\bm{v}}=v_{c,\max}/2\sim0.75$, and then $\tan\alpha\sim0.5$ and $\tan\beta\sim0.5$.
Therefore, we obtain the estimate $F_d\sim2.3$ and $F_l\sim4.5$.
The inertial force is approximated as $F_m\sim\frac{\pi}{4}K\chi_\rho\chi_h\chi_c\sin\theta\left(\dv{v}{t}-\delta V\omega\cos\theta\right)$, where we omitted the term $\propto \sin^2\theta$
because the tail amplitude is small.
In the bracket, $\dv{v}{t}$ is roughly estimated as
the change of $v$ during the half period $1/2f_a$ from Fig.~\ref{StVorb}:
$\dv{v}{t}\sim\frac{1}{2}\frac{2v_{c,\max}}{1/2f_a}\sim7.5$.
We also estimate $\omega\sim v/\chi_c\sim2$ at $\theta=0$.
Thus, we obtain $F_m\sim2.8$.
To summarize, we obtain the relation of force magnitudes $F_D\sim F_d\lesssim F_m<F_l$.
If we use two dimensional plate ($K\rightarrow1$),
the relation between $F_d,~F_l,~F_m$ 
becomes $F_d<F_l<F_m$,
as was confirmed
in the previous two-hinge two-dimensional flapping airfoil model~\cite{Nagai1996}.

%%%%%%%%%%%%%%%%%%%%%%%%%%%%%%%%%%%%%%%%%%%%%%%%%%%%%%%%%%%%
\section{Probability distributions}
\label{Appenddistri}

\begin{figure}[!t]
\centering
\includegraphics[width=0.95\linewidth]{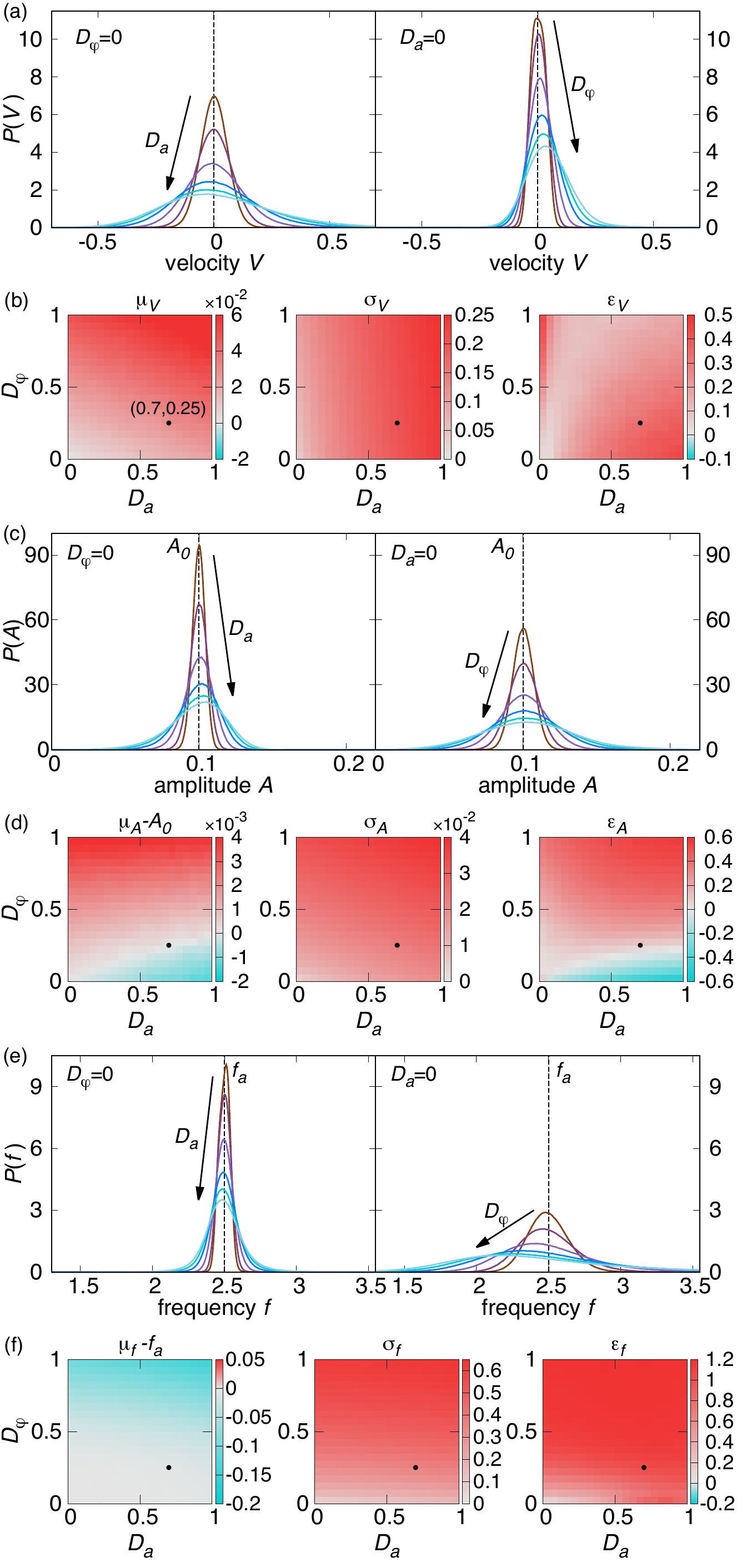}
%bb=0 0 360 252
\caption{
Dependence of the distributions on the noise strengths 
for $\chi_c=0.375,~C_\Gamma=2.0$, and $f_a=2.5$.
The distributions and statistical measures
of (a)-(b) the thrust velocity $V$, (c)-(d) tailbeat amplitude $A$, and (e)-(f) tailbeat frequency $f$.
In (a), (c), (e), the left subplot shows the dependence of the distribution
on $D_a$ with $D_\varphi=0$, and the right subplot is the dependence on $D_\varphi$ with $D_a=0$.
The values of $D_a$ or $D_\varphi$ are $\{0.05,0.1,0.25,0.5,0.75,1\}$.
Larger values of $D_a$ or $D_\varphi$ give a wider distribution.
In (b), (d), (f), the left colormap shows the expected value as a function of $D_a,~D_\varphi$,
the middle is the standard deviation, and the right is the skewness.
The filled circles correspond to $D_a=0.7$ and $D_\varphi=0.25$,
which are mainly the values used in the simulations.
}
\label{Appenddisfig}
\end{figure}

Here, we show the dependence of the distributions
on $D_a$ and $D_\varphi$ in detail.
Figs.~\ref{Appenddisfig}(a), (c), and (e) show
the probability distributions of the thrust velocity $V$, tailbeat amplitude $A$, 
and frequency $f$, respectively,
as functions $D_a$ or $D_\varphi$ (with the increment $\Delta D_a=\Delta D_\varphi=0.05$).
Qualitatively, we find that
$P(V)$ is controlled by $D_a$ rather than $D_\varphi$,
while $P(A)$ and $P(f)$ are mainly determined by $D_\varphi$.
Quantitatively,
we define the following statistical measures for $Q=V,A,f$
by using the probability distribution $P(Q)$~\cite{Joanes1998}:
the expected value 
\begin{equation}
\label{muQ}
\mu_Q=\int\dd QP(Q)Q,
\end{equation}
the variance 
\begin{equation}
\sigma_Q^2=\int\dd QP(Q)(Q-\mu_Q)^2,
\end{equation}
where $\sigma_Q$ is the standard deviation,
and the skewness
\begin{equation}
\varepsilon_Q=\frac{1}{\sigma_Q^3}\int\dd QP(Q)(Q-\mu_Q)^3.
\end{equation}
If $\varepsilon_Q>0$, 
the probability distribution has a fat tail at larger $Q$ 
and vice versa.

The average thrust velocity $\mu_V$ is less than 5\% of the background flow $U$,
but the positiveness of $\mu_V$ indicates that noises reduce the thrust speed
(see Fig.~\ref{Appenddisfig}(b));
note that $V>0$ means that the thrust speed is less than $U$.
The standard deviation $\sigma_V$ depends mainly on $D_a$ as expected,
and the skewness $\varepsilon_V$ is positive.

For the tailbeat amplitude, 
the deviation from the target amplitude 
$\mu_A-A_0$ is less than 5\% of $A_0$
as shown in Fig.~\ref{Appenddisfig}(d).
The noise strengths $D_a$ and $D_\varphi$
tend to decrease and increase the amplitude $\mu_A$, respectively.
Therefore we can tune the noise strengths continuously
so that $\mu_A\simeq A_0$ is always satisfied.
On the other hand, for the standard deviation of $A$, 
there is little difference between its dependences on $D_a$ and $D_\varphi$.
As for the skewness $\varepsilon_A$, $D_\varphi$ mainly determines its sign,
while $D_a$ controls its magnitude.

Fig.~\ref{Appenddisfig}(f) shows the statistical measures for the tailbeat frequency $f$.
We find a strong dependence of $P(f)$ on $D_\varphi$ rather than on $D_a$.
The deviation from the input frequency $\mu_f-f_a$ is always negative 
and is less than 8\% of $f_a$, and $\sigma_f$ mainly depends on $D_\varphi$ as expected.
The skewness $\varepsilon_f$ is positive.

We adopted $D_a=0.7$ and $D_\varphi=0.25$
by comparing the distributions with the experimental results
(Supplementary Information of Ref.~\cite{Li2020}).
Figs.~\ref{tevdis}(b)-(d) show the distributions for these values, which correspond to the
filled circles in Figs.~\ref{Appenddisfig}(b), (d), and (f).
The expected values $\mu_V,~\mu_A,~\mu_f$ are close to
the values $0,~A_0,~f_a$ for the noiseless case, respectively:
$\mu_V$ is less than 2\% of $U$,
$\mu_A-A_0$ is 0.03\% of $A_0$, and
$\mu_f-f_a$ is less than 0.15\% of $f_a$.
Our resolution of the noise parameters is such that
%In particular, $\mu_A$ is close to $A_0$
$\mu_A$ changes only 0.1\%-0.3\% of $A_0$
by each step increment $\Delta D_a$ and $\Delta D_\varphi$.
On the other hand, $\mu_V$ changes 0.1\%-0.2\% of $U$,
and $\mu_f$ changes 0.1\%-0.2\%  of $f_a$
by $\Delta D_a$ and  $\Delta D_\varphi$.
The height and width of the distributions $P(V),~P(A),~P(f)$
are qualitatively in good agreement with that of goldfish.
Furthermore, the asymmetry of $P(V)$ and $P(f)$ are similar to 
the experimental results: $P(V)$ has a slightly fat tail on the right sight of the peak,
while $P(f)$ has a noticeably fat tail at larger $f$.

%%%%%%%%%%%%%%%%%%%%%%%%%%%%%%%%%%%%%%%%%%%%%%%%%%%%%%%%%%%%
\section{Energy dissipation rate}
\label{Appenddissipation}

\begin{figure}[!t]
\centering
\includegraphics[width=\linewidth]{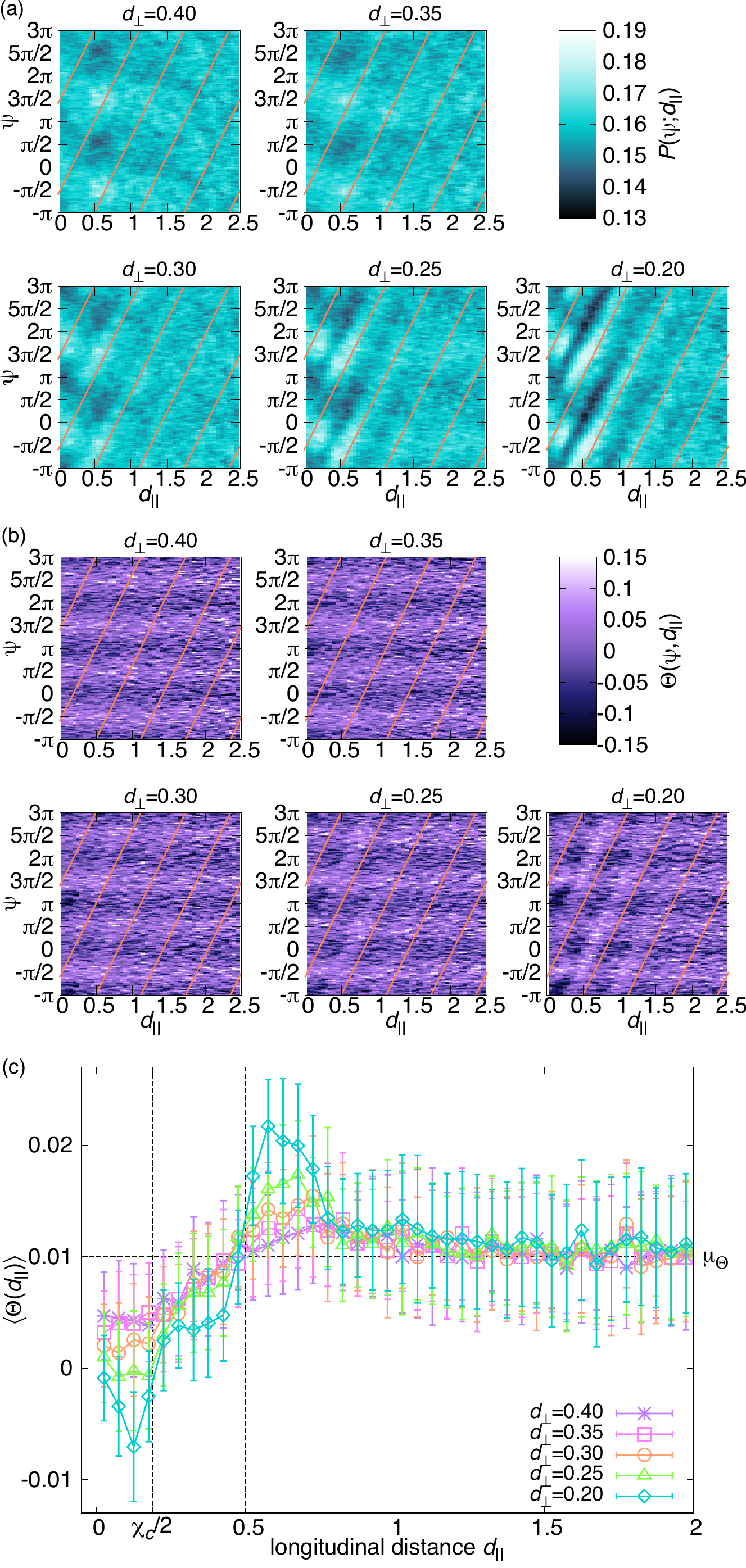}
%bb=0 0 360 252
\caption{
Dependence of (a) the probability distribution $P(\psi;d_\parallel)$,
(b) the dissipation rate $\Theta(\psi,d_\parallel)$, and
(c) $\langle\Theta(d_\perp)\rangle$ on $d_\perp$ with $C_\Gamma=2.0$.
The display scheme is the same as in Figs.~\ref{PpsiTHTCGam}(a)-(b) and Fig.~\ref{THTaveCGam}.
}
\label{PpsiTHTdperp}
\end{figure}

Here we discuss some properties of 
the energy dissipation rate for  pair swimming.
In the main text, we fixed the the lateral distance $d_\perp =0.2$.
Here we
show the dependence of $P(\psi;d_\parallel)$ on $d_\perp$ in
Fig.~\ref{PpsiTHTdperp}(a).
We find that $P(\psi;d_\parallel)$ approaches a uniform distribution
as $d_\perp$ increases to $d_\perp=0.4$.
The plot of the energy dissipation rate $\Theta(\psi,d_\parallel)$ 
in Fig.~\ref{PpsiTHTdperp}(b)
shows only the horizontal stripe pattern for $d_\perp=0.4$, 
and the oblique stripes gradually disappeared with increasing $d_\perp$.
Also, the expected value $\langle\Theta(d_\parallel)\rangle$ 
deviates negatively from $\mu_\Theta$ at $d_\parallel \lesssim 0.5$ 
for any value of $d_\perp$ (see Fig.~\ref{PpsiTHTdperp}(c)).
This energy gain is larger than the energy consumption 
in the range $d_\parallel\gtrsim0.5$.

%%%%%%%%%%%%%%%%%%%%%%%%%%%%%%%%%%%%%%%%%%%%%%%%%%%%%%%%%%%
%\section{Horizontal pattern of the dissipation rate}
%\label{Appenddissipation}

Next we provide an explanation of  the horizontal stripe pattern 
of the dissipation rate $\Theta(\psi,d_\parallel)$ 
as seen in Fig.~\ref{PpsiTHTCGam}(b),
using the statistical properties of a solitary swimmer.
From Fig.~\ref{Pphitail}(a), we can approximate the phase distribution as
\begin{equation}
P(\phi)\approx\frac{1}{2\pi}+\epsilon\cos(2\phi+\gamma_P),
\end{equation}
where $\epsilon\sim\mathcal{O}(10^{-3})$ and $\gamma_P$ is a constant.
In addition, from Fig.~\ref{Pphitail}(b), the dissipation rate $\Theta(\phi)$ 
is approximated by
\begin{equation}
\Theta(\phi)\approx\Theta_0+E\cos(2\phi-\gamma_\Theta)
\end{equation}
where $\Theta_0\sim\mathcal{O}(10^{-1})$, $E\sim\mathcal{O}(1)$, and $\gamma_\Theta$ is a constant.
Using these approximations, 
a straightforward calculation gives the average dissipation rate as
\begin{equation}
\mu_\Theta=\Theta_0+\pi E\epsilon\cos(\gamma_P+\gamma_\Theta)\sim\mathcal{O}(10^{-2})>0,
\end{equation}
which corresponds to the numerically value $\mu_\Theta\simeq0.01$ 
(see Fig.~\ref{Pphitail}(b) inset).

Next, we calculate the probability distribution of the phase difference $\psi=\phi_1-\phi_2$
neglecting the hydrodynamic interaction between the two swimmers.
As the distribution does not depend on the distance,
we denote $\widetilde{P}(\psi):=P(\psi;d_\parallel)$ for simplicity.
Exploiting the symmetry between $\phi_1$ and $\phi_2$, or $\psi$ and $-\psi$, 
we rewrite the joint probability distribution of $\phi_1$ and $\phi_2$ as
\begin{eqnarray}
P(\phi_1)P(\phi_2)
&=&P(\phi_1)\frac{1}{2}(P(\phi_1+\psi)+P(\phi_1-\psi))\nonumber\\
&=:& P(\phi_1,\psi).
\end{eqnarray}
Then $\widetilde{P}(\psi)$ is obtained by integrating 
$P(\phi_1,\psi)$ over the range $\phi_1\in[-\pi,\pi]$,
which yields
\begin{equation}
\widetilde{P}(\psi)=\frac{1}{2\pi}+\pi\epsilon^2\cos2\psi.
\end{equation}
This result indicates that a horizontal stripe pattern emerges in $P(\psi;d_\parallel)$, 
but we cannot detect it in Fig.~\ref{PpsiTHTCGam}(a)
due to the smallness of $\epsilon^2\sim\mathcal{O}(10^{-6})$.

Finally, we calculate the dissipation rate $\widetilde{\Theta}(\psi):=\Theta(\psi,d_\parallel)$
in the absence of hydrodynamic interaction.
Noting that the probability distribution of $\phi_1$ for a given value of $\psi$
is given by $P(\phi_1, \psi)/\widetilde{P}(\psi)$, 
we obtain 
\begin{eqnarray}
\widetilde{\Theta}(\psi) &=& 
\int_{-\pi}^\pi\dd\phi_1\Theta(\phi_1)\frac{P(\phi_1,\psi)}{\widetilde{P}(\psi)}\nonumber\\
&=&\Theta_0+\pi E\epsilon\cos(\gamma_P+\gamma_\Theta)(1+\cos2\psi)+\mathcal{O}(\epsilon^2)\nonumber\\
&=&\mu_\Theta+\pi E\epsilon\cos(\gamma_P+\gamma_\Theta)\cos2\psi+\mathcal{O}(\epsilon^2).
\end{eqnarray}
Therefore, the energy dissipation rate $\Theta(\psi,d_\parallel)$ 
has an ${\cal O}(\epsilon)$ deviation from $\mu_\Theta$,
which is  proportional to $\cos 2\psi$.
(Note also that $\cos(\gamma_P+\gamma_\Theta)<0$.)
This explains the horizontal stripe pattern in the plots in Fig.~\ref{PpsiTHTCGam}(b).

In addition, the expected value of $\widetilde{\Theta}(\psi)$ 
is
\begin{equation}
\langle\Theta\rangle=\int_{-\pi}^\pi\dd\psi\widetilde{\Theta}(\psi)\widetilde{P}(\psi)=\mu_\Theta+\mathcal{O}(\epsilon^2),
\end{equation}
and thus is almost equal to $\mu_\Theta$, as shown in Fig.~\ref{THTaveCGam}.

%%%%%%%%%%%%%%%%%%%%%%%%%%%%%%%%%%%%%%%%%%%%%%%%%%%%%%%%%%%


\begin{thebibliography}{999}

\bibitem{Conradt2005}
L. Conradt and T. J. Roper,
Trends Ecol. Evol.
\textbf{20},
449
(2005).

\bibitem{Vicsek2012}
T. Vicsek and A. Zafeiris,
Phys. Rep.
\textbf{517},
71
(2012).

\bibitem{Parrish2002}
J. K. Parrish, S. V. Viscido, and D. Gr\"{u}nbaum,
Biol. Bull.
\textbf{202},
296
(2002).

\bibitem{Lopez2012}
U. Lopez, J. Gautrais, I. D. Couzin, and G. Theraulaz,
Interface Focus
\textbf{2},
693
(2012).

\bibitem{Terayama2015}
K. Terayama, H. Hioki, and M. Sakagami,
Int. J. Semant. Comput.
\textbf{9},
143
(2015).

\bibitem{Harpaz2020}
R. Harpaz, E. Schneidman,
eLife,
\textbf{9},
e56196
(2020).
%doi: 10.7554/eLife.56196

% 削除
%\bibitem{Rosenthal2015}
%S. B. Rosenthal, C. R. Twomey, A. T. Hartnett, H. S. Wu, and I. D. Couzin,
%Proc. Natl. Acad. Sci. U.S.A.
%\textbf{112},
%4690
%(2015).

\bibitem{Liao2007}
J. C. A. Liao,
Philos. Trans. R. Soc. B
\textbf{362},
1973
(2007).

\bibitem{Breder1954}
C. M. Breder, 
Ecology 
\textbf{35}, 
361 
(1954).

\bibitem{Niwa1994}
H. Niwa, 
J. Theor. Biol.
\textbf{171}, 
123 
(1994).

\bibitem{Aoki1982}
I. Aoki, 
Bull. Jpn. Soc. Sci. Fish.
\textbf{48}, 
1081 
(1982).

\bibitem{Huth1992}
A. Huth and C. Wissel, 
J. Theor. Biol.
\textbf{156}, 
365 
(1992).

\bibitem{Huth1994}
A. Huth and C. Wissel, 
Ecol. Model.
\textbf{75}, 
135 
(1994).


\bibitem{Couzin2002}%%%
I. D. Couzin, J. Krause, R. James, G. D. Ruxton, and N. R. Franks,
J. Theor. Biol.
\textbf{218}, 
1
(2002).

%\bibitem{DOrsogna2006}%%%
%M. R. D’Orsogna et al., Phys. Rev. Lett. 
%\textbf{96}, 
%104302
%（2006）.

%\bibitem{Nguyen2012}%%%
%N. H. P. Nguyen, E. Jankowski, and S. C. Glotzer, 
%Phys. Rev. E
%\textbf{86}, 
%011136
%(2012).

\bibitem{Gautrais2012}
J. Gautrais, F. Ginelli, R. Fournier, S. Blanco, M. Soria, H. Chat\'e, and G. Theraulaz, 
PLOS Comput. Biol.
\textbf{8}, 
e1002678 
(2012).

\bibitem{Calovi2014}
D. S. Calovi, U. Lopez, S. Ngo, C. Sire, H. Chat\'e, and G. Theraulaz,
New J. Phys. 
\textbf{16}, 
015026 
(2014).

\bibitem{Filella2018}
A. Filella, F. Nadal, C. Sire, E. Kanso, and C. Eloy, 
Phys. Rev. Lett.
\textbf{120}, 
198101 
(2018).

\bibitem{Deng2021}
J. Deng and D. Liu,
Bioinspir. Biomim.
\textbf{16},
046013
(2021).

\bibitem{Ito2022a}
S. Ito and N. Uchida,
J. Phys. Soc. Jpn.
\textbf{91}, 
064806 
(2022).

\bibitem{Ito2022b}
S. Ito and N. Uchida,
Europhys. Lett.
\textbf{138}, 
17001 
(2022).

\bibitem{Hemelrijk2008}
C. K. Hemelrijk and H. Hildenbrandt, 
\textit{Ethology}
\textbf{114},
245
(2008).

\bibitem{Bastien2020}
R. Bastien and P. A. Romanczuk,
Sci. Adv.
\textbf{6},
eaay0792
(2020).

\bibitem{Tchieu2012}
A. A. Tchieu, E. Kanso, and P. K. Newton,
Proc. R. Soc. A
\textbf{468},
3006
(2012).

\bibitem{Gazzola2016}
M. Gazzola, A. A. Tchieu, D. Alexeev, A. de Brauer, and P. Koumoutsakos,
J. Fluid Mech.
\textbf{789}, 
726 
(2016).

\bibitem{Lauder2002}
G. V. Lauder and E. G. Drucker, 
News Physiol. Sci., 
\textbf{17},
235 
(2002).

\bibitem{Akanyeti2017}
O. Akanyeti, J. Putney, Y. R. Yanagitsuru, G. V. Lauder, W. J. Stewart, and J. C. Liao,
Proc. Natl. Acad. Sci. U.S.A.
\textbf{114},
13828
(2017).

\bibitem{Wise2018}
T. N. Wise, M. A. B. Schwalbe, and E. D. Tytell, 
J. Exp. Biol.
\textbf{221}, 
jeb190892 
(2018).

\bibitem{Breder1965}
C. M. Breder,
Zoologica
\textbf{50},
97
(1965).

\bibitem{Weihs1973}
D. Weihs,
Nature
\textbf{241},
290
(1973).

%\bibitem{Badgerow1981}
%J. P. Badgerow and F. R. Hainsworth, 
%J. Theor. Biol.
%\textbf{93}, 
%41 
%(1981).

\bibitem{Partridge1979}
B. L. Partridge and T. J. Pitcher,
Nature
\textbf{279},
418
(1979).

\bibitem{Marras2014}
S. Marras, S. S. Killen, J. Lindstr\"{o}m, D. J. McKenzie, J. F. Steffensen, and P. Domenici,
Behav. Ecol. Sociobiol.
\textbf{69},
219
(2014).

\bibitem{Ashraf2016}
I. Ashraf, R. Godoy-Diana, J. Halloy, B. Collignon, and B. Thiria,
J. R. Soc. Interface
\textbf{13},
20160734
(2016).

\bibitem{Ashraf2017}
I. Ashraf, H. Bradshawa, T. Ha, J. Halloy, R. Godoy-Diana, and B. Thiria,
Proc. Natl. Acad. Sci. U.S.A.
\textbf{114},
9599
(2017).

\bibitem{Li2020}
L. Li, M. Nagy, J. M. Graving, J. Bak-Coleman, G. Xie, and I. D. Couzin,
Nat. Commun.
\textbf{11},
5408
(2020).

\bibitem{Elfring2009}%%%
G. J. Elfring and E. Lauga,
\textit{Phys. Rev. Lett.}
\textbf{103},
088101
(2009).

\bibitem{Liao2021}%%%
W. Liao and E. Lauga,
\textit{Phys. Rev. E}
\textbf{103},
042419
(2021).


\bibitem{Dewey2014}
P. A. Dewey, D. B. Quinn, B. M. Boschitsch, and A. J. Smits,
Phys. Fluids
\textbf{26},
041903
(2014).

\bibitem{Boschitsch2014}
B. M. Boschitsch, P. A. Dewey, and A. J. Smits,
Phys. Fluids
\textbf{26},
051901
(2014).

\bibitem{Becker2015}
A. D. Becker, H. Masoud, J. W. Newbolt, M. Shelley, and L. Ristroph,
Nat. Commun.
\textbf{6}, 
97
(2015).

\bibitem{Ramananarivo2016}%%%
S. Ramananarivo, F. Fang, A. Oza, J. Zhang, and L. Ristroph,
Phys. Rev. Fluids 
\textbf{1},
071201(R)
(2016).

\bibitem{Newbolt2019}
J. W. Newbolt, J. Zhang, and L. Ristroph,
Proc. Natl. Acad. Sci. U.S.A.
\textbf{46}, 
2419
(2019).

\bibitem{Oza2019}
A. U. Oza, L. Ristroph, and M. J. Shelley,
Phys. Rev. X 
\textbf{9}, 
041024
(2019).

\bibitem{Zhu2014}
X. Zhu, G. He, and X. Zhang,
Phys. Rev. Lett. 
\textbf{113}, 
238105
(2014).

\bibitem{Park2018}
S. G. Park and H. J. Sung,
J. Fluid Mech.
\textbf{840}, 
154
(2018).

\bibitem{Peng2018}
Z. Peng, H. Huang, and X. Lu,
J. Fluid Mech. 
\textbf{849}, 
1068
(2018).

\bibitem{Hemelrijk2015}
C. K. Hemelrijk, D. A. P. Reid, H. Hildenbrandt, and J. T. Padding,
Fish. Fish.
\textbf{16},
511
(2015).

\bibitem{Daghooghi2015}
M. Daghooghi and I. Borazjani,
Bioinspir. Biomim.
\textbf{10},
056018
(2015).

\bibitem{Maertens2017}
A. P. Maertens, A. Gao, and M. S. Triantafyllou,
J. Fluid Mech.
\textbf{813},
301
(2017).

\bibitem{Li2019}
G. Li, D. Kolomenskiy, H. Liu, B. Thiria, and R. Godoy-Diana,
PLOS ONE
\textbf{14},
e0215265
(2019).

\bibitem{XLi2021}
X. Li, J. Gu, Z. Su, and Z. Yao,
Phys. Fluids
\textbf{33},
121905
(2021).

\bibitem{Pan2022}
Y. Pan and H. Dong,
Phys. Fluids
\textbf{34},
111902
(2022).

\bibitem{Kelly2023}
J. Kelly, P. Yu, A. Menzer, and D. Haibo,
Phys. Fluids
\textbf{35},
041906
(2023).

\bibitem{Lin2023}
Z. Lin, D. Liang, A. P. S. Bhalla, A. A. S. Al-Shabab, M. Skote, W. Zheng, and Y. Zhang,
Phys. Fluids
\textbf{35},
081901
(2023).

\bibitem{Sfakiotakis1999}
M. Sfakiotakis, D. M. Lane, and J. B. C. Davies, 
IEEE J. Ocean. Eng.
\textbf{24},
237
(1999).

\bibitem{Lauder2005}
G. V. Lauder and E. D. Tytell, 
Fish Physiol.
\textbf{23}, 
425 
(2005).

\bibitem{Azuma1992}
A. Azuma,
\textit{The Biokinetics of Flying and Swimming} 
(Springer, Tokyo, 1992).

\bibitem{Nagai1996}
M. Nagai, I. Teruya, K. Uechi, and T. Miyazato,
Trans. Jpn. Soc. Mech. Eng. B
\textbf{62},
200
(1996).
(\url{https://www.jstage.jst.go.jp/article/kikaib1979/62/593/62_593_200/_pdf/-char/en})

\bibitem{Hirayama2000}
M. Hirayama, T. Nagamatsu, and K. Ueda,
Mem. Fac. Fish. Kagoshima Univ.
\textbf{49},
17
(2000).
(\url{https://ir.kagoshima-u.ac.jp/?action=repository_uri&item_id=6287&file_id=16&file_no=1})

\bibitem{Gazzola2015}
M. Gazzola, M. Argentina, and L. Mahadevan,
Proc. Natl. Acad. Sci. U.S.A.
\textbf{112},
3874
(2015).


\bibitem{Jones1999}
R. E. Jones, R. J. Petrell, and D. Pauly,
Aquac. Eng. 
\textbf{20}, 
216
(1999).

\bibitem{Landau1987}
L. D. Landau and E. M. Lifshitz,
\textit{Fluid Mechanics}
(Butterworth-Heinemann, New York, 1987).

\bibitem{Bainbridge1958}
R. Bainbridge,
J. Exp. Biol.
\textbf{35},
109
(1958).

\bibitem{Hunter1971}
J. R. Hunter and J. R. Zweifel
Fish. Bull.
\textbf{69},
253
(1971).

\bibitem{Webb1984}
P. W. Webb, P. T. Kostecki, and E. Don Stevens,
J. Exp. Biol.
\textbf{109},
77
(1984).

\bibitem{Li2021}
G. Li, H. Liu, U. K. M\"{u}ller, C. J. Voesenek, and J. L. van Leeuwen,
Proc. R. Soc. B
\textbf{288},
20211601
(2021).

\bibitem{NoteA0}
Note that, in the context of experiments of
Refs.~\cite{Bainbridge1958,Hunter1971,Webb1984,Akanyeti2017,Li2021},
they measure the peak to peak amplitude $2A_0$ as ``the amplitude".

\bibitem{Taylor1952}
G. I. Taylor,
Proc. Roy. Soc. Lond. A
\textbf{214},
158
(1952).

\bibitem{Lighthill1960}
M. J. Lighthill,
J. Fluid Mech.
\textbf{9},
305
(1960).

\bibitem{Landau1986}
L. D. Landau, E. M. Lifshitz, A. M. Kosevich, and L. P. Pitaevskii,
\textit{Theory of Elasticity}
(Butterworth-Heinemann, New York, 1986).

\bibitem{McHenry1995}
M. J. McHenry, C. A. Pell, and J. H. Long Jr,
J. Exp. Biol.
\textbf{198},
2293
(1995).

\bibitem{Grillner2006}
S. Grillner,
Neuron
\textbf{52},
751
(2006).

\bibitem{Song2020}
J. Song, I. Pallucchi, J. Ausborn, K. Ampatzis, M. Bertuzzi, P. Fontanel, L. D. Picton, and A. El Manira,
Neuron
\textbf{105},
1048
(2020).

\bibitem{Ekeberg1999}
\"{O}. Ekeberg and S. Grillner,
Phil. Trans. R. Soc. Lond. B
\textbf{354},
895
(1999).

\bibitem{Matsuoka2011}
K. Matsuoka,
Biol. Cybern.
\textbf{104},
297
(2011).

\bibitem{Schwalbe2019}
M. A. B. Schwalbe, A. L. Boden, T. N. Wise, and E. D. Tytell,
Sci. Rep.
\textbf{9},
8088
(2019).

%\bibitem{Uhlenbeck1930}
%G. E. Uhlenbeck and L. S. Ornstein,
%Phys. Rev.
%\textbf{36},
%823
%(1930).

\bibitem{Giaiotti2006}
D. B. Giaiotti and F. Stel, 
\textit{The Rankine vortex model}, 
University of Trieste-International Centre for Theoretical Physics
(2006).
(\url{https://moodle2.units.it/pluginfile.php/109093/mod\_resource/content/1/rankine-vortex-notes.pdf})

\bibitem{Drucker1999}
E. G. Drucker and G. V. Lauder,
J. Exp. Biol. 
\textbf{202}, 
2393 
(1999).

\bibitem{Schnipper2009}
T. Schnipper, A. Andersen, and T. Bohr,
J. Fluid Mech.
\textbf{633},
411
(2009).

\bibitem{Agre2016}
N. Agre, S. Childress, J. Zhang, and L. Ristroph,
Phys. Rev. Fluids
\textbf{1},
033202
(2016)

\bibitem{Ortiz2015}
X. Ortiz, D. Rival, and D. Wood,
Energies
\textbf{8},
2438
(2015).

\bibitem{Brennen1982}
C. E. Brennen,
\textit{A Review of Added Mass and Fluid Inertial Forces}
(Naval Civil Engineering Laboratory, Port Hueneme, 1982).

%\bibitem{McKean1969}
%H. P. McKean,
%\textit{Stochastic Integrals}
%(Academic Press, New York, 1969).

\bibitem{Tanaka1996}
I. Tanaka and M. Nagai,
\textit{Teikou to suishin no ryuutai rikigaku:
suisei seibutsu no kousoku yuuei nouryoku ni manabu},
(Ship \& Ocean Foundation, Tokyo, 1996).
(\url{https://www.spf.org/_opri_media/publication/pdf/199609_rp040220.pdf})

\bibitem{Nagai1979}
M. Nagai,
Nagare
\textbf{10},
47
(1979).
(\url{https://www.jstage.jst.go.jp/article/nagare1970/10/4/10_4_47/_pdf/-char/en})

\bibitem{Bainbridge1963}
R. Bainbridge,
J. Exp. Biol.
\textbf{40},
23
(1963).

\bibitem{Gazzola2014}
M. Gazzola, M. Argentina, and L. Mahadevan,
Nat. Phys.
\textbf{10},
758
(2014).

\bibitem{Triantafyllou1993}
G. S. Triantafyllou, M. S. Triantafyllou, and M. A. Grosenbaugh,
J. Fluids Struct.
\textbf{7}, 
205 
(1993).

\bibitem{Taylor2003}
G. K. Taylor, R. L. Nudds, and A. L. R. Thomas,
Nature
\textbf{425},
707
(2003).

\bibitem{Domenici1997}
P. Domenici and R. W. Blake,
\textit{J. Exp. Biol.}
\textbf{200},
1165
(1997).

\bibitem{GLi2021}
G. Li, I. Ashraf, B. Fran\c{c}ois, D. Kolomenskiy, F. Lechenault, R. Godoy-Diana, and B. Thiria,
\textit{Commun. Biol.}
\textbf{4},
40
(2021).


\bibitem{Shadwick2008}
R. E. Shadwick and D. A. Syme,
\textit{J. Exp. Biol.}
\textbf{211},
1603
(2008).

\bibitem{Santo2021}
V. Di Santo, E. Goerig, D. K. Wainwright, O. Akanyeti, J. C. Liao, T. Castro-Santos, and G. V. Lauder, 
Proc. Natl. Acad. Sci. U.S.A.
\textbf{118},
e2113206118
(2021).


\bibitem{Wen2013}
L. Wen and G. Lauder, 
Bioinspir. Biomim.
\textbf{8},
046013
(2013).

\bibitem{Lucas2020}
K. N. Lucas, G. V. Lauder, and E. D. Tytell, 
Proc. Natl. Acad. Sci. U.S.A.
\textbf{117},
10585
(2020).

\bibitem{Thandiackal2023}
R. Thandiackal and G. Lauder,
eLife
\textbf{12},
e81392
(2023).
 
\bibitem{Nauen2002}
J. C. Nauen and G. V. Lauder,
J. Exp. Biol.
\textbf{205},
1709
(2002).


%Appendix
\bibitem{Huang2014}
N. E. Huang and S. S. P. Shen,
\textit{Hilbert-Huang Transform and Its Applications}
(World Scientific, Singapore, 2014).

\bibitem{Jiang2014}
H. Jiang, Y. Li, and Z. Cheng,
Appl. Mech. Mater.
\textbf{518},
161
(2014).

\bibitem{Wendel1956}
K. Wendel,
\textit{Hydrodynamic Masses and Hydrodynamic Moments of Inertia}
(MIT Libraries, Cambridge, 1956).

\bibitem{Ilio2018}
G. Di Ilio, D. Chiappini, S. Ubertini, G. Bella, S. Succi,
Comput. Fluids
\textbf{166},
200
(2018).

\bibitem{Tandler2019}
T. Tandler, E. Gellman, D. De La Cruz, and D. J. Ellerby,
J. Fish Biol.
\textbf{94},
532
(2019).

\bibitem{Wu1977}
T. Y. Wu, 
in \textit{Scale Effects in Animal Locomotion}, 
ed. T. J. Pedley 
(Academic Press, New York, 1977) 
pp. 203.

\bibitem{Joanes1998}
D. N. Joanes and C. A. Gill,
\textit{J. R. Stat. Soc. (Ser. D)}
\textbf{47},
183
(1998).



\end{thebibliography}
\end{document}